\documentclass[prx,amsmath,amssymb, notitlepage, twocolumn,nofootinbib,superscriptaddress
]{revtex4-1}
\usepackage[colorlinks=true,citecolor=blue,linkcolor=blue]{hyperref}

\usepackage{graphicx}
\usepackage{color}
\usepackage{dcolumn}
\usepackage{bm}
\usepackage{mathrsfs}
\usepackage{amsmath}
\usepackage{amssymb}
\usepackage{amsthm}
\usepackage{amsfonts}
\usepackage{enumerate}
\usepackage{latexsym}
\usepackage{hyperref}
\usepackage{centernot}
\usepackage{multirow}

\begin{document}

\title{Efficient Topological Materials Discovery Using Symmetry Indicators}

\begin{abstract}
Although the richness of spatial symmetries has led to a rapidly expanding inventory of possible topological crystalline (TC) phases of electrons, physical realizations have been slow to materialize due to the practical difficulty to ascertaining band topology in realistic calculations.
Here, we integrate the recently established theory of symmetry indicators of band topology into first-principle band-structure calculations, and test it on a databases of previously synthesized crystals. The combined algorithm is found to efficiently unearth topological materials and predict topological properties like protected surface states.
On applying our algorithm to just 8 out of the 230 space groups, we already discover numerous materials candidates displaying a  diversity of topological phenomena, which are simultaneously captured in a single sweep.
The list includes recently proposed classes of TC insulators that had no previous materials realization as well as other topological phases, including:
(i) a screw-protected 3D TC insulator,  $\beta $-MoTe$_{2}$, with gapped surfaces except for 1D  helical ``hinge'' states;
(ii) a rotation-protected TC insulator $ {\rm BiBr}$ with coexisting surface Dirac cones and hinge states;
(iii) non-centrosymmetric $\mathbb Z_2$ topological insulators undetectable using the well-established parity criterion, AgXO (X=Na,K,Rb);
(iv) a Dirac semimetal MgBi$_{2}$O$_{6}$;
(v) a Dirac nodal-line semimetal AgF$_2$; and
(vi) a metal with three-fold degenerate band crossing near the Fermi energy, AuLiMgSn.
Our work showcases how the recent theoretical insights on the fundamentals of band structures can aid in the practical goal of discovering new topological materials.
\end{abstract}

\date{\today}
\author{Feng Tang}
\affiliation{National Laboratory of Solid State Microstructures and School of Physics, Nanjing University, Nanjing 210093, China}
\affiliation{Collaborative Innovation Center of Advanced Microstructures, Nanjing 210093, China}

\author{Hoi Chun Po }
\affiliation{Department of Physics, Harvard University, Cambridge, MA 02138, USA}

 \author{Ashvin Vishwanath}
 \affiliation{Department of Physics, Harvard University, Cambridge, MA 02138, USA}

\author{Xiangang Wan}
\affiliation{National Laboratory of Solid State Microstructures and School of Physics, Nanjing University, Nanjing 210093, China}
\affiliation{Collaborative Innovation Center of Advanced Microstructures, Nanjing 210093, China}

\maketitle

\section{Introduction}
Topological materials, exemplified by the topological insulators (TIs) and nodal semimetals, showcase intriguing physical properties which could not emerge had electrons been classical point-like particles \cite{review-1,review-2, SM-RMP}.
Most of the known varieties of topological phenomena arising in weakly correlated materials require symmetry protection, and the inherent richness of spatial symmetries of crystals manifests as a corresponding diversity of distinct electronic phases. Depending on the spatial symmetries at play, an insulating 3D topological material could be characterized by surface states featuring protected conical or quadratic dispersions \cite{TCI-1,TCI-2,TCI-3,wallpaper}, interesting band connectivities \cite{hourglass}, or even 1D gapless modes pinned to the hinges of the crystal facets \cite{HOTI,FangFu,TCI-4,Fang-5,TCI-6, Fang-3, Khalaf}. Moreover, spatial symmetries also stabilize band crossings and give rise to various kinds of nodal semimetals \cite{SM-RMP}.

There is much interest in finding concrete materials candidates for the various topological crystalline (TC) phases. A conventional ``designer'' approach can be summarized as follows: first, one specifies a topological (crystalline) phase of interest, and, typically through intuition, identifies possible materials classes which have the approach symmetry and chemistry ingredients; second, one performs realistic first-principle electronic calculations; third, one extracts the band topology from the calculation, either through the evaluation of wave-function-based invariants \cite{review-1,review-2} or, when applicable, through simple criteria relating symmetry representations to specific forms of band topology \cite{Fu-Kane, Fang-1, Ari, Hughes11}.
Such criteria are exemplified by the celebrated Fu-Kane parity criterion for $\mathbb Z_2$ TIs \cite{Fu-Kane}. They offer immense computational advantage in the diagnosis of topological materials, as their evaluation only requires physical data defined on a small set of isolated momenta.

In the described conventional approach, the analysis starts with a particular topological phase in mind. Correspondingly, a negative result only implies the targeted form of band topology is absent, but does not rule out the existence of other forms of nontrivial band topology.
With the ever-expanding catalogue of TC phases, such oversight of band topology is nearly inevitable, and one may have to re-analyze all the previously studied materials in the search of a newly proposed class of topological (crystalline) phase of matter.

Here, we propose an alternative paradigm for the discovery of topological materials which can help overcoming the mentioned challenge.
Instead of hunting for materials candidates with a targeted form of band topology in mind, our framework automatically singles out all the materials featuring any form of robust band topology, provided that it is detectable using symmetry representations\footnote{By ``robust'' here, we refer in particular to band topology that is stable against the addition of trivial degrees of freedom, which includes all the usual topological phases with protected surface states. ``Fragile topology,'' as introduced in Ref.\ \onlinecite{Fragile}, is not robust in this sense, despite the fact that they can be detectable from symmetry representations. Such fragile phases will not be captured in the materials search algorithm described in this work.}.
Our approach is inspired by the recent wave of development on establishing a more comprehensive understanding between symmetry representations and band topology  \cite{wallpaper-Kane, Po, QCT, Po-2, Fang-3, Fang-4, Khalaf}. In particular, our algorithm is built upon the exhaustive theory of symmetry indicators (SIs) developed in Ref.\ \cite{Po}, which enables a reliable discovery of topological materials even without a priori knowledge on the precise form of the band topology involved.
Remarkably, the same algorithm can be applied to evaluate the SI of any materials in any space group, which leads to a versatile method well-suited for handling the diversity of TC phases.

To demonstrate the power of this paradigm, we apply our program and search for materials reported in a database of previously synthesized compounds \cite{icsd}, focusing on 8 of the 230 space groups. We discover topological materials featuring an array of distinct topological properties, ranging from weak and strong topological insulators to the recently proposed ``higher-order'' topological crystalline insulators (TCIs) \cite{HOTI,FangFu,TCI-4,Fang-5,TCI-6, Fang-3, Khalaf}. In addition, our algorithm also captures topological semimetals. In the following, we will highlight four particular materials candidates discovered: (i) $\beta$-MoTe$_2$, a screw-protected TCI with helical 1D hinge states ; (ii) BiBr, a $C_2$-rotation protected TCI \cite{FangFu} with  2D gapless Dirac-cone surface states coexisting with 1D hinge states on the side; (iii) AgNaO, a non-centrosymmetric strong TI detected using a new SI utilizing the improper four-fold rotation $S_4$ \cite{Fang-3, Khalaf}; and (iv) Bi$_2$MgO$_6$, a Dirac semimetal with a clean Fermi surface consisting only of two symmetry-related Dirac points.
Other materials found, including weak TI, TCI with hourglass surface states, semimetal with triply degenerate nodal point, and nodal-line semimetal, are described in the supplementary materials (SM) \cite{SM}.

\section{Materials diagnosis through symmetry indicators
\label{sec:SI-theory}}
We will begin by briefly reviewing the theory of symmetry indicators in Ref.\ \onlinecite{Po}, which is rested upon the observation that, insofar as symmetry representations are concerned, one can represent any set of bands of interest by a vector\footnote{Technically, these integer-valued ``vectors'' are not truly vectors; we will nonetheless abuse the terminology here to highlight the linear structure they obey.} \cite{Ari, wallpaper-Kane}:
\begin{eqnarray}
\mathrm{\mathbf{n}} &=&(\nu ,n_{\mathbf{k_{1}}}^{1},n_{\mathbf{k}%
_{1}}^{2},\ldots n_{\mathbf{k}_{1}}^{\alpha _{1}},\ldots ,n_{\mathbf{k}%
_{2}}^{1},n_{\mathbf{k_{2}}}^{2},\ldots ,  \notag \\
&&n_{\mathbf{k}_{2}}^{\alpha _{2}},\ldots ,n_{\mathbf{k_{N}}}^{1},n_{\mathbf{%
k_{N}}}^{2},\ldots ,n_{\mathbf{k}_{N}}^{\alpha _{N}},\dots ),  \label{bs}
\end{eqnarray}%
where $\nu $ is the total number of the filled energy bands. The subscript $%
\mathbf{k}_{1},\mathbf{k_{2}},\ldots ,\mathbf{k}_{N}$ runs over the distinct classes of high-symmetry momenta in the Brillouin zone (BZ), and the superscript $1,2,\ldots
,\alpha _{i},\ldots $ labels the irreducible
representation (irrep) of the little group  ($\mathcal{G}_{\mathbf{k}_{i}}$) at $\mathbf{k}_{i}$.
Here, $n_{\mathbf{k}_{i}}^{\alpha _{i}}$ denotes the
number of times an irrep $\alpha _{i}$ of $\mathcal{G}_{\mathbf{k}_{i}}$
appears within the bands of interest, and should be integer-valued if the bands are separated from above and below by band gaps at all the momenta $\mathbf k_{i}$.
However, if the Fermi energy intersects with the energy bands at a momentum $\mathbf{k}_s$, then the values of $n_{\mathbf{k}_{s}}^{\alpha _{s}}$ are ill-defined. Nonetheless, one can still formally compute $n_{\mathbf{k}_{s}}^{\alpha _{s}}$ using standard methods, and the valued obtained are generically not integer-valued.

While Eq.\ \eqref{bs} is defined for a general set of bands, there are strong constraints on $\mathbf{n}$ if the bands correspond to a trivial atomic insulator (AI), i.e., one can construct a full set of symmetry-respecting Wannier orbitals, whose centers fall into one of the Wyckoff positions and transform under symmetries according to the representations of the corresponding site-symmetry groups.
Conversely,  any combination of Wyckoff positions with a choice on the site-symmetry group representations gives rise to an AI. The data on Wyckoff positions and their associated irreducible symmetry representations (irreps) have been exhaustively tabulated \cite{Bradley, ITC, Bilbao}. This allows one to readily construct all the possible vectors $\mathbf{n}$ corresponding to an AI (see an example in the supplemental material (SM) \cite{SM}).

As any AI is viewed as a vector, it can be expanded over a basis. More concretely, from the mentioned computation one can obtain $d_{%
\mathrm{AI}}$ basis vectors $ \{ \mathbf{a}_{i}~:~i=1,2,\ldots ,d_{\mathrm{AI}}\}$, such that for any vector $\mathbf{n}_{\rm AI}$ arising from an AI, one can expand
\begin{equation}
\mathrm{\mathbf{n}}_{\rm AI}=\sum_{i=1}^{d_{\mathrm{AI}}} m_{i}\mathbf{a}_{i},
\label{bsonai}
\end{equation}%
where $m_i$ are integers.
The entries appearing in the basis vectors $\mathbf{a}_{i}$ are all integer-valued, and we denote the largest common factor of the entries by $C_i$. As, like any vector space, the basis vectors are not uniquely defined, one may wonder if $C_i$ has any physical meaning. However, there exist special choices of basis for which the values of $C_i$ are maximized, and such basis can be found through the Smith normal decomposition.
As concrete examples, the  AI basis vectors we use in this work, chosen to maximize $C_i$, are all listed in the SM \cite{SM}.
Unlike the basis vectors themselves, the set of maximized common factors $\{ C_1, \dots C_{d_{\rm AI}}\}$ is fixed for any symmetry setting, and one may further label the basis vectors such that $C_1 \leq C_2\leq \dots \leq C_{d_{\rm AI}}$.

While our discussion so far is restricted to the symmetry analysis of trivial atomic insulators, it forms the anchor for an efficient diagnosis of topological materials achieved by integrating the described with first-principle calculations.
To see how, consider a material to be diagnosed. We first perform a routine band-structure calculation to obtain $n_{\mathbf{k}}^{\alpha }$, the multiplicities of the symmetry representations at the high-symmetry momenta $\mathbf k_i$ for the filled bands.
We then subject $\mathbf{n}$ to the same expansion of the AI basis $\mathbf{a}_i$ as in Eq.\ \eqref{bsonai} to arrive at a  collection of expansion coefficients $\{ q_i\}$.
 If the materials of interest are time-reversal invariant and features significant spin-orbit coupling, the coefficients $\{ q_i\}$ can be classified into the following three cases (the discussion below has to be appropriately modified for other symmetry settings \cite{Po-2, Fang-4}):\\

\noindent \textbf{Case 1} A band gap is found at all momenta $\mathbf{k}_i$ and ${q_{i}}$'s are all integers.

This indicates that $\mathrm{\mathbf{n}}$ is formally indistinguishable from an AI as far as (stable) symmetry representations are concerned.
As such, the material at hand is either an atomic insulator, an accidental (semi-)metal, or possesses band topology that is either indiscernible from a representation viewpoint \cite{Fang-1, Po, Khalaf} or is not stable against the addition of trivial degrees of freedom \cite{Po,Fragile}. A more refined wave-function-based analysis is required if a more comprehensive understanding on the material is desired.\\

\begin{table}[!tbp]
\caption{We focus on the following space groups ($\mathcal{SG}$s), in which a strong topological insulators generate a $\mathbb Z_4$ subgroup  in the group of symmetry indicators, $X_{\mathrm{BS}}$. The entry $2\in \mathbb Z_4$ corresponds to various kinds of topological crystalline insulators, and the predicted materials candidates for such phases  are tabulated.
}
\label{tab:sg}\centering
\par
\begin{tabular}{|c|c|c|c|c|}
\hline
$X_{\mathrm{BS}}$ & $\mathbb{Z}_{2}^3\times\mathbb{Z}_{4}$ & $\mathbb{Z}_{2}^2\times
\mathbb{Z}_{4}$ & $\mathbb{Z}_{2}\times \mathbb{Z}_{4}$ & $\mathbb{Z}_{4}$
\\ \hline
$\mathcal{SG}$ & $\mathbf{2}$ & $\mathbf{11}$,$\mathbf{12}$ & $\mathbf{166}$
& $\mathbf{61,136,227}$ \\ \hline
Materials & Ag$_2$F$_5$ & $\beta$-MoTe$_2$,BiBr & A7-P & c-TiS$_2$ \\ \hline
\end{tabular}%
\end{table}

\noindent \textbf{Case 2}: A band gap is found at all momenta $\mathbf{k}_i$ but some $q_i$'s are not integers. Nonetheless, ${q_{i}\mathrm{C}_{i}}$'s are all integers.

Barring the possibility of additional accidental band crossings, which one can readily check by computing the energy (but not wave-functions) of the bands over the entire BZ, the material at hand is a band insulator with nontrivial band topology \cite{Po, Fang-3, Khalaf}. Furthermore, the topological properties of the material can be exposed by evaluating the symmetry indicator (SI) $ r_i \equiv {q_{i}\mathrm{C}_{i}} \mod \mathrm{C}_{i}$, which takes value in $0, 1,\dots, \mathrm{C}_{i}-1$.
Note that, if $C_i=1$, then $r_i$ is necessarily $0$ and conveys no information. As we have arranged the integers $C_i$ in ascending order, we can assume $C_i =1$ for all $i < p$ and hence drop $r_1,\dots, r_{p-1}$ from the discussion.
The SI can then be viewed as an element of the group \cite{Po} $X_{\rm BS} \equiv \mathbb Z_{\mathrm{C}_p}\times \mathbb Z_{\mathrm{C}_{p+1}}\times \dots \times \mathbb Z_{\mathrm{C}_{d_{\rm AI}}}$, and one can infer the possible forms of band topology associated to the materials  from the results in Refs.\ \onlinecite{Fang-3, Khalaf}.
Note that, if $X_{\rm BS}$ is trivial for a space group, then this case can never occur.

We remark that, as we will see below, for some cases there could be distinct forms of band topology giving rise to the same SI, and as such one has to further evaluate their associated band invariants to pinpoint its concrete topological properties.\\

\noindent \textbf{Case 3}: Not all of ${q_{i}\mathrm{C}_{i}}^{\prime }s$ are integers.

As proven in Ref.\ \onlinecite{Po}, this indicates that the material is a symmetry-protected (semi-)metal, where a (nodal) Fermi surface is unavoidable given the symmetries and filling of the crystal. Note that this case can occur even when $X_{\rm BS}$ is trivial.

From the discussion above, we see that cases 2 and 3 respectively correspond to promising materials candidates for topological crystalline insulators and (semi-)metals, with the only uncertainty being the detailed energetics which might bring about unwanted trivial bands near the Fermi energy.
Therefore, by screening materials through computing their SIs, one can channel most of the computational effort to these promising candidates for an efficient discovery of new topological material.

\begin{figure}[tbp]
\includegraphics[width= 0.4 \textwidth]{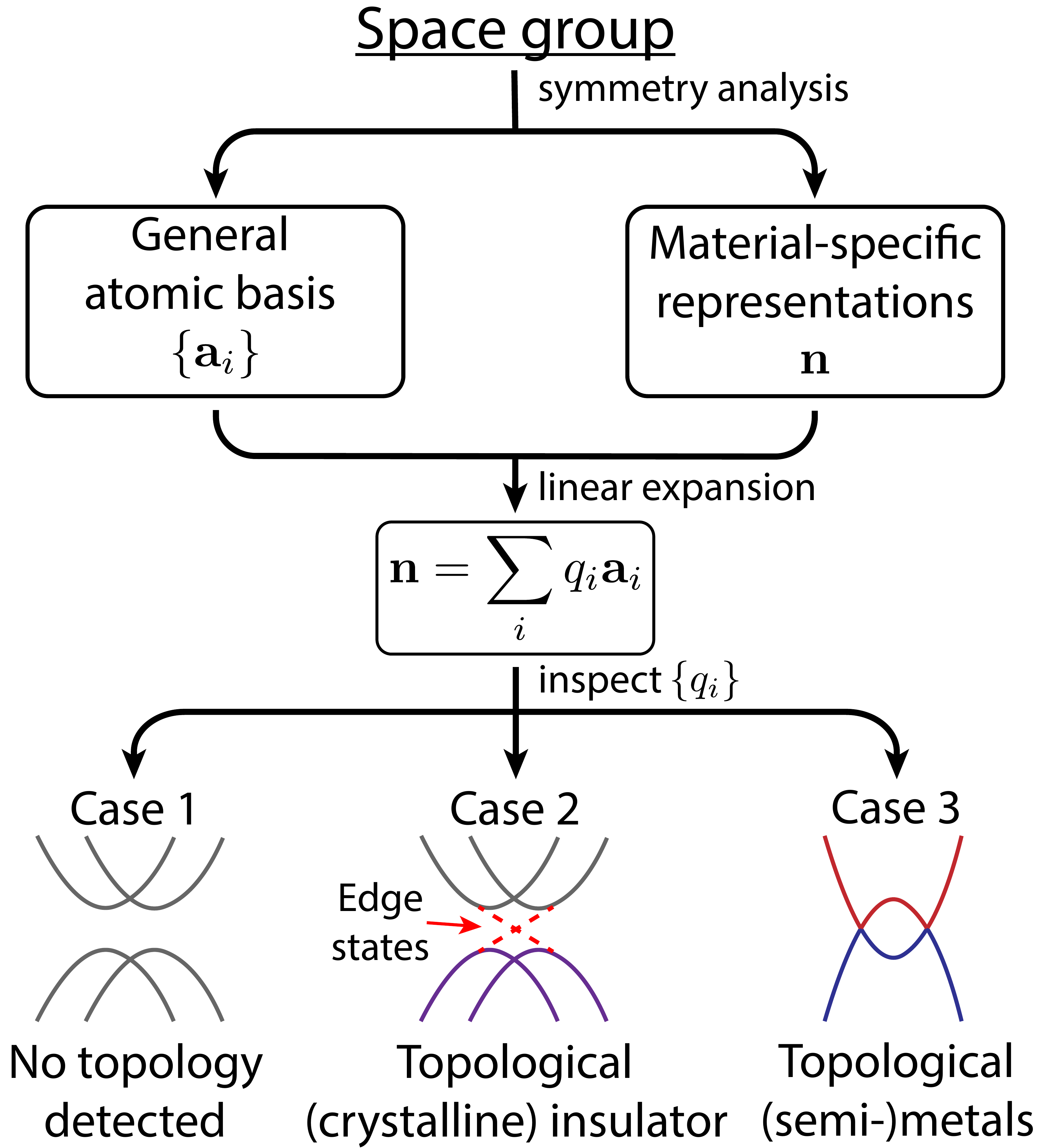}\newline
\caption{Topological materials discovery algorithm. The expansion coefficients $\{ q_i \} $ of the materials-specific representation vector $\mathbf{n}$ against the atomic basis $\{ \mathbf{a}_i\}$ can be classified into three cases. Cases 2 and 3 correspond respectively to promising materials candidates for topological (crystalline) insulators and semimetals.}
\label{fig:flow}
\end{figure}

In the following, we apply our search strategy (Fig. \ref{fig:flow}) to discover a variety of topological materials. In particular, we will focus on experimentally synthesized crystals \cite{icsd} with significant spin-orbit coupling and no magnetic atoms. Furthermore, we restrict our attention to materials with $\leq 30$ atoms in the primitive unit cell.

We will focus our search to a number of space groups for which the SI group $X_{\rm BS}$ is particularly interesting. First, we will focus on centrosymmetric space groups (i.e., containing inversion), for which $X_{\rm BS}$  (in the present symmetry setting) contains a factor of $\mathbb Z_4$.
The  generator $1 \in \mathbb Z_4$ corresponds to a strong TI, and the $\mathbb Z_4$ structure, as discussed in Ref.\ \onlinecite{Po}, implies that taking two copies of an inversion-symmetric strong TI results in a nontrivial insulator even when all the conventional $\mathbb Z_2$ TI indices have been trivialized. It was later realized that such materials with SI $2\in \mathbb Z_4$ will always possess symmetry-protected surface states \cite{FangFu, Fang-3, Khalaf}, and they showcase a very rich phenomenology which depends on the precise symmetry setting at hand: the sruface states could be 2D in nature when protected by a mirror, glide, or a rotation symmetry, or may manifest as 1D hinge states when protected by inversion or screw symmetries (Fig.\ \ref{fig:MoTe}a). We will provide explicit materials candidates for TCIs with SI $2 \in \mathbb Z_4$, which are discovered through our search among materials crystalizing in the centrosymmetric space groups ($\mathcal{SG}$s) $\mathbf{2}$, $\mathbf{11}$, $\mathbf{12}$, $\mathbf{61}$, $\mathbf{136}$, $\mathbf{166}$, $\mathbf{227}$. In particular, note that these space groups have $X_{\rm BS} = (\mathbb Z_2)^{j} \times \mathbb Z_4$ with $j=0,1,2,3$ (Table \ref{tab:sg}), where the $\mathbb Z_2$ factors correspond to either weak TIs or weak mirror Chern insulators.
In addition, we further consider the non-centrosymmetric space group $\mathbf{216}$, which contains the improper four-fold rotation $S_4$ symmetry, which enables one to diagnose a strong TI through SI \cite{Po, Fang-3, Khalaf} even when the Fu-Kane criterion \cite{Fu-Kane} is not applicable.

Aside from TCIs corresponding to case 2 above, we also encountered numerous materials realizing case 3 above, which correspond to symmetry-protected (semi-)metals. We will also discuss the particularly promising candidates found.

\section{Screw-protected hinge-states in $\beta$-M\lowercase{o}T\lowercase{e}$_2$}
Our first example is the transition-metal dichalcogenide (TMD) $\beta$-MoTe$_{2}$, discovered through a search of materials crystalizing in space group ($\mathcal{SG}$) $\mathbf{11}$ ($P2_{1}/m$).
MoTe$_{2}$ has three different structural phases: hexagonal $\alpha $-phase \cite{mote2-alpha}, monoclinic $\beta $-phase \cite{mote2-beta} and
orthorhombic $\gamma $-phase \cite{mote2-gamma}. At room temperature, $\beta
$-MoTe$_{2}$ forms a monoclinic structure with $\mathcal{SG}\mathbf{11}$ \cite{mote2-beta}. Its monolayer crystal has been proposed as a candidate of quantum spin Hall insulator
\cite{mote2-qsh}. Furthermore, around 240--260 K it undergoes a
structural transition to the non-centrosymmetric $\gamma $-phase \cite{mote2-strtra}. The $\gamma $-phase was theoretically predicted to be a
type-II Weyl semimetal  and has been verified experimentally \cite{mote2-wsm-1,mote2-wsm-2,mote2-wsm-3,mote2-wsm-4}.
Interestingly, we found that $\beta$-MoTe$_{2}$ in bulk crystal form is a $2\in \mathbb Z_4$ TCI in $\mathcal{SG}\mathbf{11}$, which features 1D helical hinge surface states protected by a two-fold screw symmetry.

\begin{figure}[htbp]
\centering
\includegraphics[width=0.55 \textwidth]{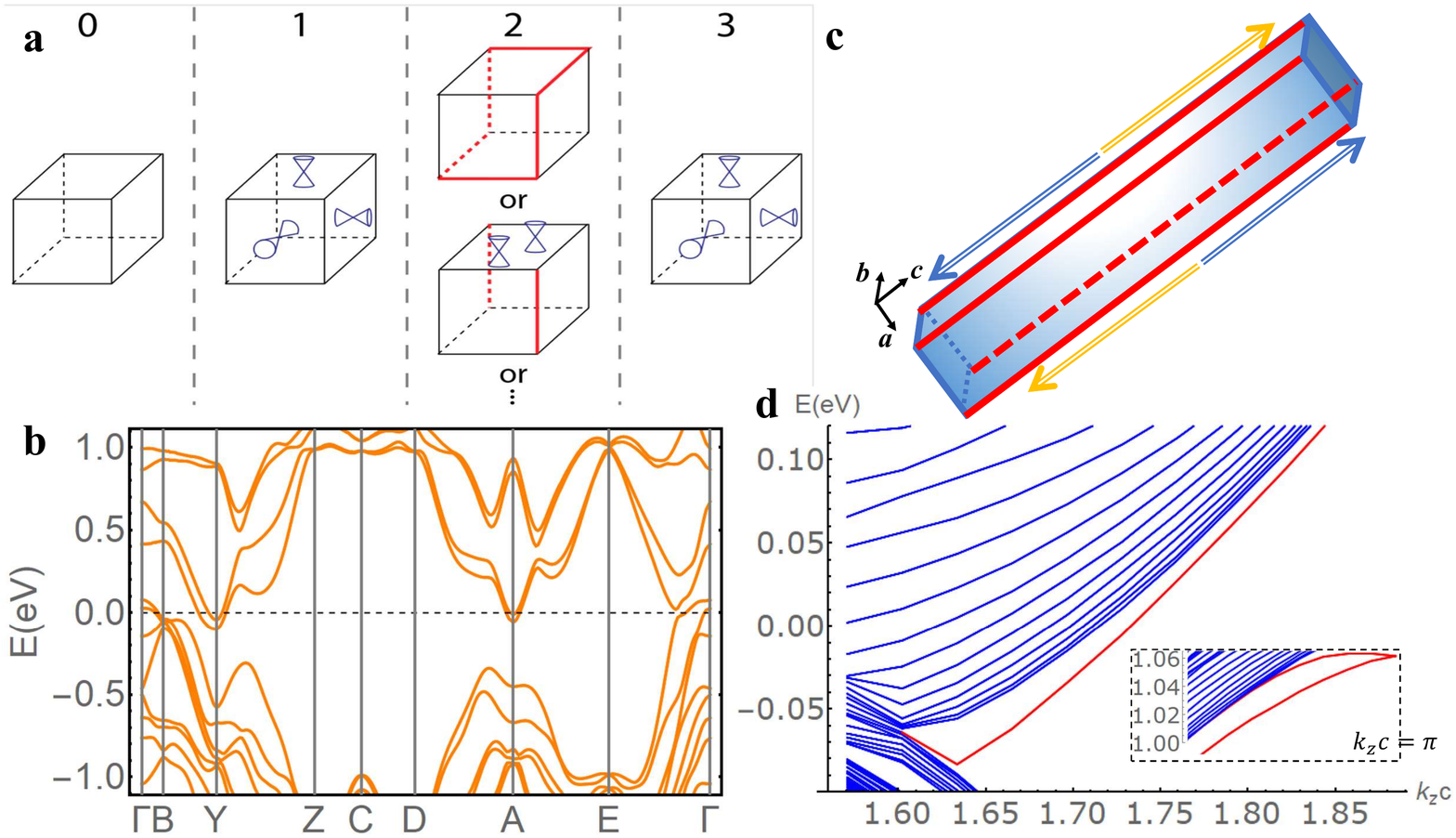}
\caption{
(a) Representative topological (crystalline) phases corresponding to the $\mathbb Z_4$ subgroup of the symmetry indicators considered.
While materials with indicators $1,3\in \mathbb Z_4$ correspond to conventional strong topological insulators, and that $0 \in \mathbb Z_4$ is consistent with a trivial phase, systems with indicator $2\in \mathbb Z_2$ can correspond to a diverse set of topological phases protected by crystalline symmetries. It is worth mentioning  that topological phase transition among these phases can be obtained through parity switch (details are given in SM \cite{SM}).
(b) The electronic structure of bulk $\protect\beta$-MoTe$_{2}$
from first principles calculation.
(c) Geometric setting for the calculation in (d), where open boundary conditions along the a- and b-axes with lengths of $20$ and $50$ lattice constants respectively, and periodic boundary condition is imposed along the screw axis (i.e., $c$-axis).
(d) The electronic structure on the prism depicted in (c). The red lines correspond to the dispersions of the hinge states (each is two-fold degenerate).  Inset shows the same plot but for a different energy window, showing in particular the expected Kramers degeneracy at $k_z = \pi/c$.
We find there are four bands whose eigenstates are mainly localized at  two of the four hinges between the (100) and (010) planes, as indicated by thick red lines in (c).
These hinge states showcase a characteristic spin-momentum locking indicating their helical nature, as expected from the theoretical predictions \cite{Fang-3, Khalaf}. This is shown schematically by the colored arrows in (c), which denote to the direction of motion for the opposite spins.}
\label{fig:MoTe}
\end{figure}

As shown in Fig. \ref{fig:MoTe}b, in spite of the presence of electron
and hole pockets, $\beta $-MoTe$_{2}$\ has finite direct band gap throughout
the whole BZ.
For $\mathcal{SG}\mathbf{11}$, there are $d_{\rm AI}=5$ AI basis vectors \cite{Po}, which we denote by $\mathbf{a}_i$ for $i=1,\dots, 5$.
Using the convention described in Sec.\ \ref{sec:SI-theory}, one finds $C_1 = C_2 = 1$,  $C_3=C_4=2$, and $C_5=4$. This implies the SI group is $X_{\rm BS} = \mathbb Z_2 \times \mathbb Z_2 \times \mathbb Z_4$ (details are shown in the SM \cite{SM}).
Based on the electronic structure calculation, we calculate the irrep multiplicities ${n_{\mathbf{k}}^{\alpha }}^{\prime }s$ for
the $56$ valence\ bands, and obtain the representation vector  \textbf{n}. Expanding
\textbf{n}\  as in Eq. \ref{bsonai}, we get $%
(q_{1},q_{2},q_{3},q_{4},q_{5})=(12,2,1,1,\frac{1}{2})$. Thus,  $\beta $-MoTe$_{2}$\ has a SI of $(0,0,2) \in \mathbb Z_2 \times \mathbb Z_2 \times \mathbb Z_4$, which indicates non-trivial band topology.

As one can verify explicitly through the inversion eigenvalues of the bands, the SI $(0,0,2)$ for $\mathcal{SG}\mathbf{11}$ implies the conventional $\mathbb Z_2$ TI invariants are all trivial \cite{Po}. Rather, the entry $2\in \mathbb Z_4$ can be interpreted through a recently introduced inversion topological invariant $\kappa _{1}$ \cite{Khalaf,Fang-3}.
The SI alone, however, does not uniquely determine the precise form of band topology for $\mathcal{SG}\mathbf{11}$ \cite{Fang-3, Khalaf}, since it can correspond to either a mirror-protected TCI characterized by a nontrivial mirror Chern number (MCN), or a $2_1$ screw-protected TCI with characteristic hinge-surface states.
These two scenarios can be further distinguished by computing the MCN on the two planes with $k_z = 0$ and $k_z = \pi/c$. Our results, computed through the WIEN2k package \cite{wien2k}, show that the MCNs vanish for both planes (see SM for details \cite{SM}). Therefore, we conclude $\beta$-MoTe$_2$ is a candidate for the $2_1$ screw-protected TCI.

To verify the above theoretical predictions, we study the electronic structure on an inversion
symmetric  prism as shown in Fig. \ref{fig:MoTe}c. We impose open boundary conditions along two directions, and periodic boundary condition along the screw axis.
Due to the large size of the supercell, we compute the band structure through a tight binding (TB) model  by considering orthogonal Mo's $d$ orbitals and Te's $p$
orbitals (details are shown in the SM \cite{SM}). Not only owning the same
topological feature (i.e. the same ${n_{\mathbf{k}}^\alpha}'s$, $\kappa_{1}$ and MCNs), the TB model
also reasonably reproduces the bulk electronic structure of $\beta $-MoTe$_{2}$ \cite{SM}. The electronic structure on the prism is shown in Fig. \ref{fig:MoTe}d.
From the spatial profile of the eigenstates, we identity four modes mainly
localized at two of the hinges between (100) and
(010) planes (Fig. \ref{fig:MoTe}c), which correspond  to the helical 1D hinge states.

\section{$C_{2}$-protected Dirac surface states and coexisting hinge states in B\lowercase{i}B\lowercase{r}}
Next, we consider the monoclinic $\mathcal{SG} \mathbf{12}$ ($C2/m$), which contains a (symmorphic) $C_2$ rotation symmetry.
Unlike the $2_1$ screw, which is not respected on any surface, $C_2$ is respected on the surface normal to the rotation axis, and hence one expects 2D gapless surface states on for $C_2$-protected topological phases. Furthermore, it was found that such 2D surface states coexist with 1D hinge states \cite{FangFu}.

\begin{figure}[htbp]
\centering
\includegraphics[width=0.55 \textwidth]{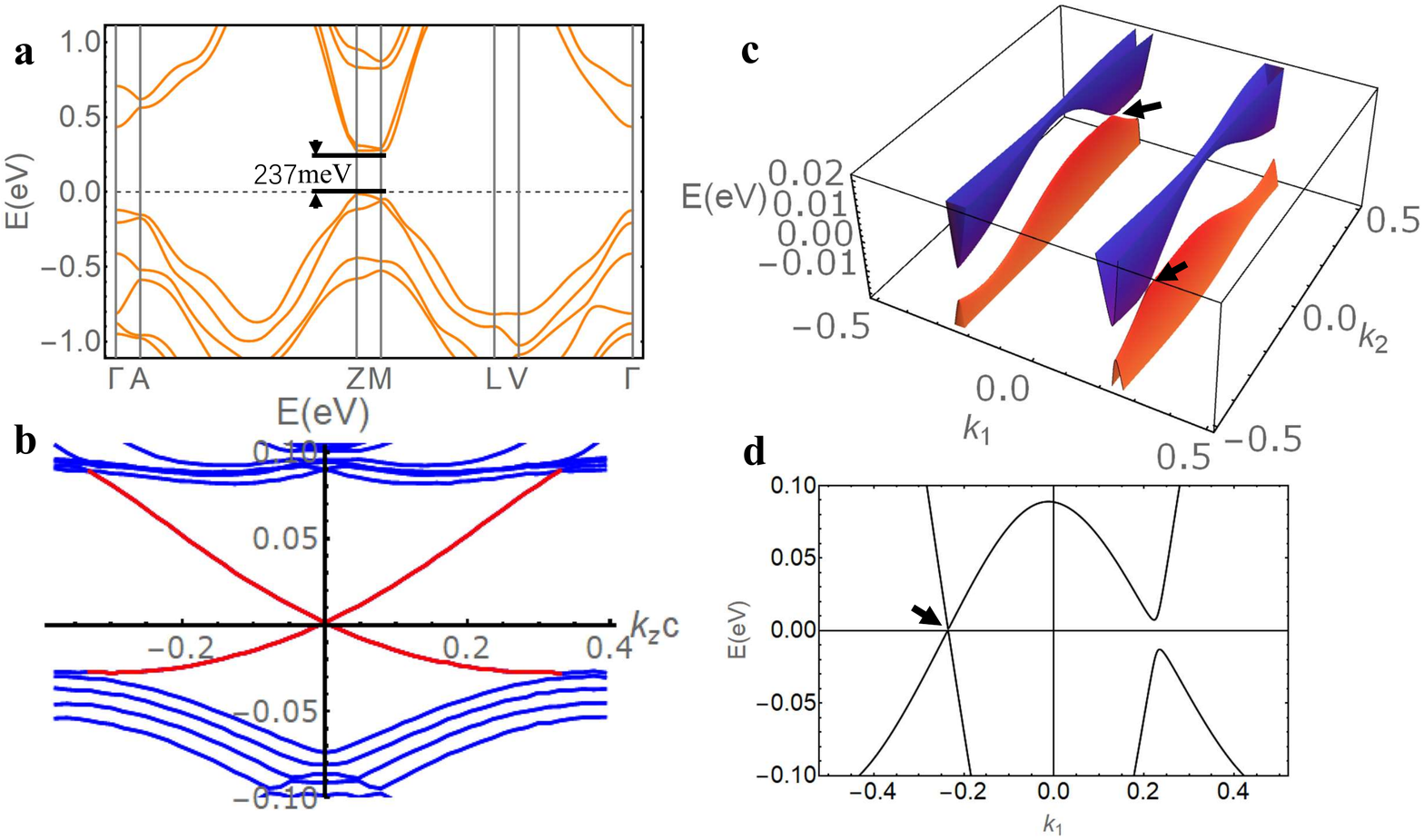}
\caption{
(a) Electronic structure of bulk BiBr from density
functional calculation.
(b) The hinge states along $\mathbf{c}$
direction of a centrosymmetric prism for BiBr; the red lines denote  hinge states traversing the bulk gap. Similar to the $\beta$-MoTe$_2$ case, they are mainly localized at opposite hinges related by $C_2$ rotation.
(c)  (001)-surface states for BiBr based on a tight-binding model. Two symmetry-protected Dirac cones are found, as indicated by the black arrows.
(d) A line cut of (b) at $k_{2}=0.333$, which clearly shows the Dirac band
crossing at about $k_{1}=-0.237$.
}
\label{fig:BiBr}
\end{figure}

Similar to the previous discussion, we first compute the AI basis vectors, from which we recover $d_{\rm AI} = 7$ and  the SI group $X_{\mathrm{BS}}^{\mathcal{SG}%
12}=\mathbb{Z}_{2}\times \mathbb{Z}_{2}\times \mathbb{Z}_{4}$ \cite{Po,SM}. This is then used in the materials search, and we identify BiBr \cite{bibr} as a TCI with a SI of $(0,0,2)$.
This compound is a insulator with a considerably large band gap ($\sim 0.24$ eV), as shown
in Fig. \ref{fig:BiBr}a \cite{SM}. This SI guarantees that the conventional $\mathbb Z_2$ TI indices all vanish \cite{Po}, and it remains to determine if the band topology is protected by mirror or the $C_2$ rotation symmetry \cite{Fang-3,Khalaf}.
To this end, we computed the MCNs for the mirror planes $k_{z}=0$ and $k_{z}=\frac{\pi }{c}$,
and found that they both vanish.
This establishes BiBr as a first example of the $C_2$-rotation protected TCI.

To further analyze the properties of BiBr, we construct a TB model starting with the  $p$ orbitals for both Bi and Br  \cite{SM}. Reproducing the electronic
structures well, our TB model is also topologically equivalent to that of
the first principles calculation results. To verify the presence of hinge states, we take an
inversion-symmetric prism geometry similar to Fig.\ \ref{fig:MoTe}c, with periodic boundary condition along the $(0,0,1)$ direction (i.e. $k_{z}$ is conserved)  but open boundary conditions for the $(0,1,0)$ and $(1/2,0,1/2)$ directions (the
directions are denoted using the conventional lattice basis vectors \cite{SM}).
The band structure of the TB model on the prism is shown in Fig. \ref{fig:BiBr}(b). As expected, a pair of helical hinge states are found.

In addition, two symmetry-related Dirac cones predicted on the $(0,0,1)$ surface of this TCI \cite{FangFu}.
This is verified explicitly by computing the electronic structure in a slab geometry. The results are shown in Fig.\ \ref{fig:BiBr}(c,d), which clearly demonstrate the existence
of the $C_{2}$-protected surface Dirac cones.
On each surface, there are two Dirac cones located
at the generic, but $C_2$-related, pair of points ($(k_{1}^{D},k_{2}^{D})=\pm (-0.237,0.333)$ measured with respect to the surface reciprocal lattice vectors.

\section{Non-centrosymmetric strong TI A\lowercase{g}N\lowercase{a}O}
While the Fu-Kane parity criterion \cite{Fu-Kane} is instrumental in the discovery of $\mathbb Z_2$ TIs, it is not applicable to crystals without inversion symmetry.
Since a Weyl semimetallic phase is expected to intervene a trivial-topological phase transition in the absence of inversion symmetry \cite{Murakami}, it is also of great interest to discover TI materials classes in non-centrosymmetric $\mathcal{SG}$s.
Curiously, unlike the conventional approach where the inapplicability of the Fu-Kane criterion demands a wave-function-based computation of the $\mathbb Z_2$ invariants \cite{Kane,Yao}, it was found that the improper four-fold rotation $S_4$ allows one to diagnose the strong TI index using SI \cite{Fang-3, Khalaf}.

\begin{figure}[htbp!]
\includegraphics[width=7.5cm]{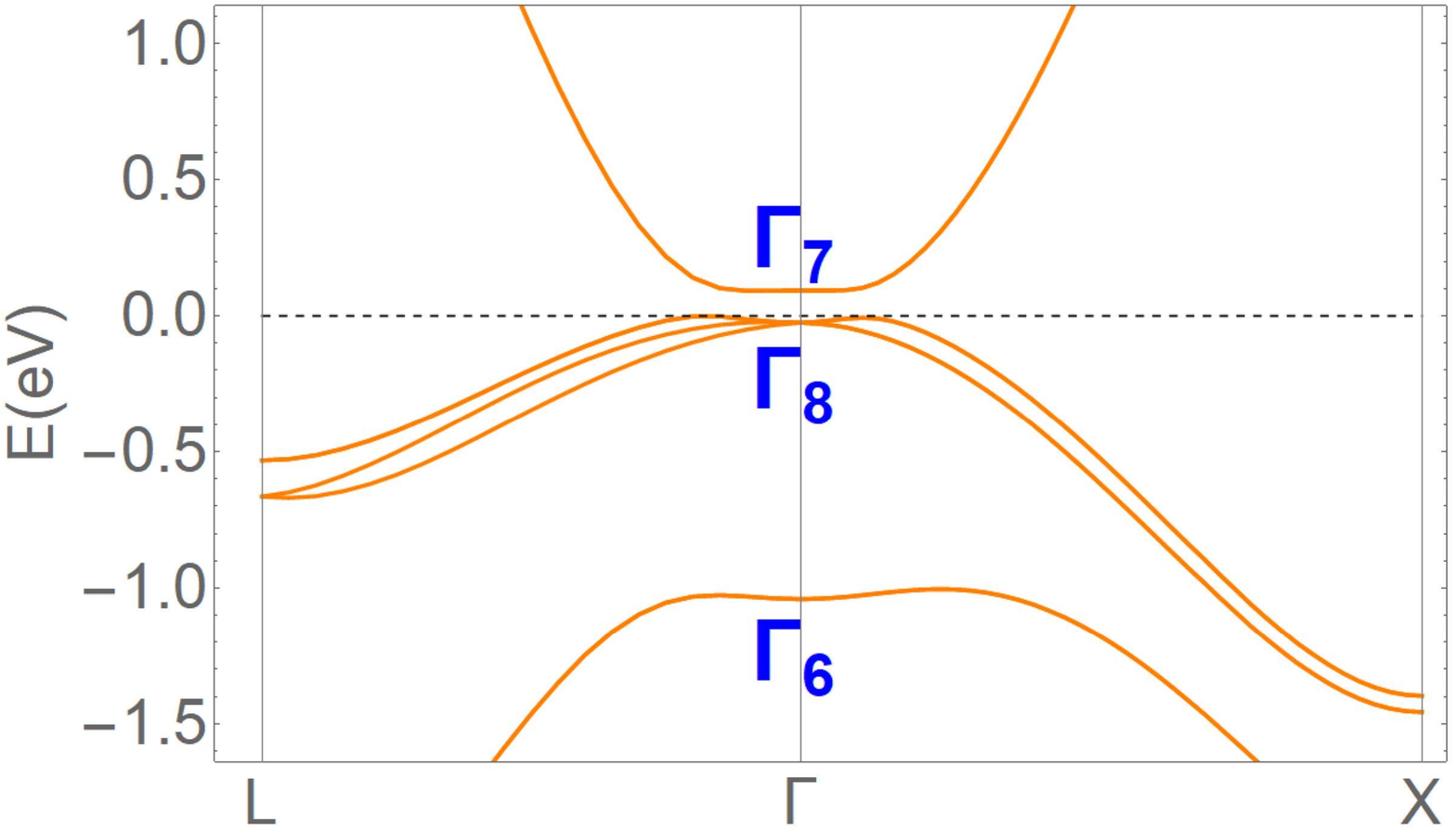}\newline
\caption{Electronic structure of the non-centrosymmetric strong topological insulator AgNaO. $\Gamma _{6,7}$ and $%
\Gamma _{8}$ label the two 2D irreps and one 4D irrep, respectively. }
\label{fig:AgNaO}
\end{figure}

In view of this, we apply our search algorithm to the $\mathcal{SG}$$\mathbf{216}$ ($F\bar 4 3 m$), which are non-centrosymmetric and contains $S_4$ in the point group. We find AgXO (X=Na,K,Rb) \cite{agnao} to be insulators with the SI $ 1 \in \mathbb{Z}_{2}$.
In the following, we take AgNaO as an example, with the corresponding discussions for AgKO  and AgRbO relegated to the SM \cite{SM}.
As shown in Fig. \ref{fig:AgNaO}, AgNaO has  a full gap ($\sim 83meV$).
Its strong TI nature can also be understood from its orbital characters:
throughout the BZ, the $s$ ($d$) band is mainly above (below) the Fermi
level. However, there is a band inversion near the $\Gamma$ point, where  the $s$-like $\Gamma _{6}$ bands are below the $d$-like $\Gamma_{7}$ bands (Fig.\  \ref{fig:AgNaO})  . As shown in the SM \cite{SM}, this band-inversion pattern results in an $S_4$ invariant, as defined in Refs.\ \onlinecite{Fang-3, Khalaf}, of $\kappa_4=1$, which is consistent with the claim that it is a strong TI.

\section{Dirac semimetal M\lowercase{g}B\lowercase{i}$_{2}$O$_{6}$}
As illustrated in Sec.\ \ref{sec:SI-theory} and highlighted in Fig.\ \ref{fig:flow}, our algorithm is not limited to discovery topological (crystalline) insulators, but is also well-suited to unearthing topological semimetals. Here, we report our discovery that MgBi$_{2}$O$_{6}$ is a Dirac semimetal with a nodal Fermi surface comprising only of a pair of symmetry-related Dirac points.

MgBi$_{2}$O$_{6}$ crystalizes in the centrosymmetric $\mathcal{SG}$$\mathbf{136}$ ($P4_2/mnm$) \cite{Bi2MgO6}. We found that its expansion coefficients $\{ q_i \}$ fall into case 3 of Fig.\ \ref{fig:flow}, which dictates necessarily gaplessness at the Fermi surface. Further band structure calculation reveals that the symmetry-dictated band crossings locate exactly at the Fermi energy without any accompanying trivial bands.
The stability of the band crossing points can be understood as follows: near the Fermi level, the low energy electronic behavior is mainly determined by Bi-$6s $ states (conduction bands) and O-$2p$ states (valence bands). Along the high-symmetry line $\Gamma$-Z, the Bi-$6s $ and O-$2p$ respectively furnish the $\Lambda_6$ and $\Lambda_7$ representations of the little group $C_{4v}$. In addition, there is a band inversion ($\sim 0.2$eV) at the $\Gamma $ point, and therefore a symmetry-protected band crossing occurs along $\Gamma$-Z. In the SM \cite{SM}, we further show from a $\mathbf{k}\cdot \mathbf{p}$ analysis that the band crossing leads to a Dirac point.

\begin{figure}[htbp!]
\includegraphics[width=7.5cm]{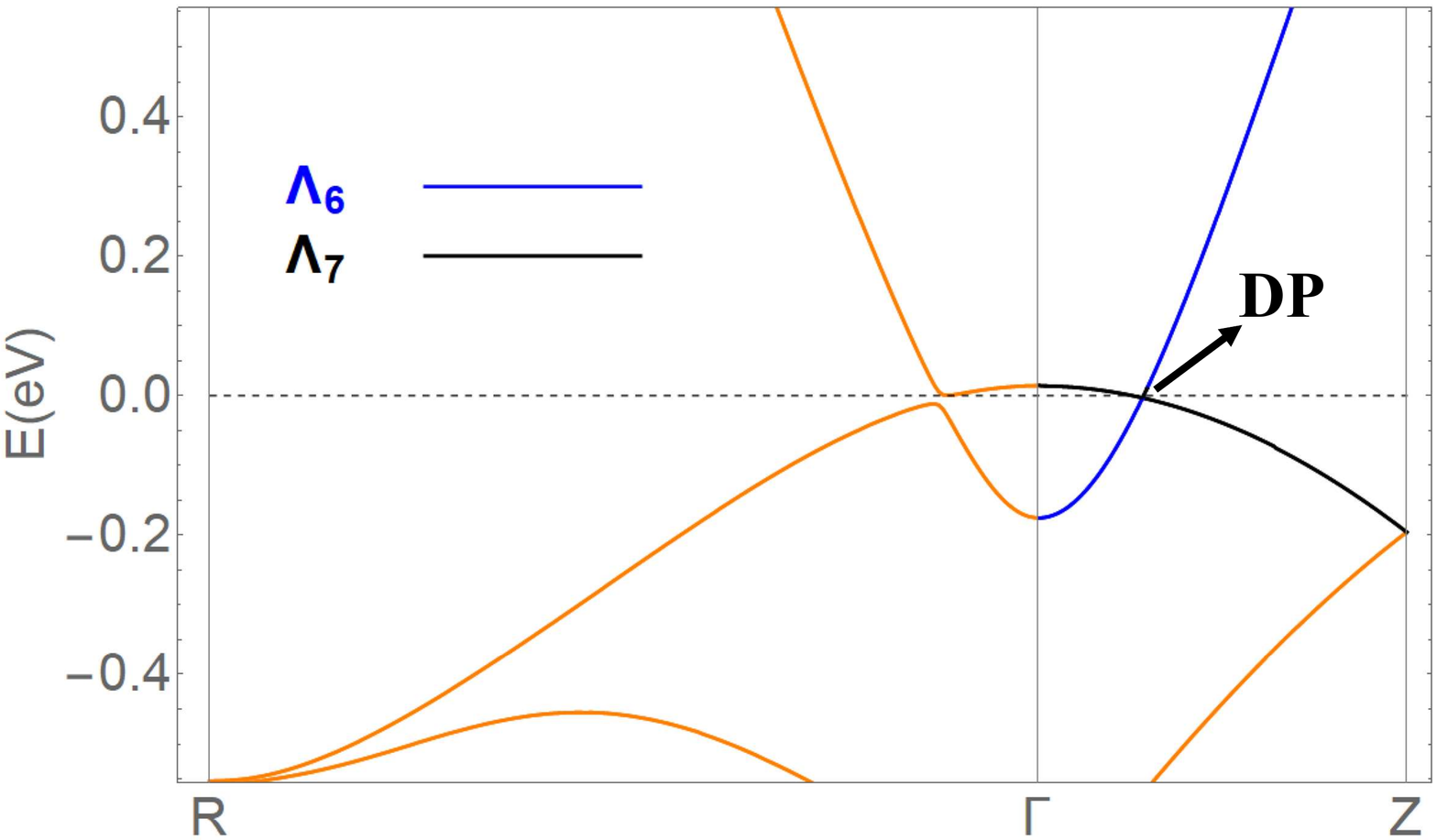}\newline
\caption{Electronic structure of Dirac semimetal MgBi$_{2}$O$_{6}$. Along
the $\Gamma$- Z line, there is a Dirac point originated from the crossing of two different 2D irreps ($\Lambda_{6}$ and $\Lambda_{7}$, indicated by black and blue lines
respectively) which is protected by $C_{4v}$ symmetry.}
\label{fig:d}
\end{figure}

\section{Other topological materials discovered}
\textit{Other $2 \in \mathbb{Z}_{4}$ materials.} Besides $\beta $-MoTe$_{2}$ and BiBr, we also find three other materials candidates with a SI of $2 \in \mathbb{Z}_{4} \leq X_{\rm BS}$.
We will summarize the findings below, and relegate the detailed analysis to the SM \cite{SM}.

First, we found that the cubic crystal TiS$_{2}$ \cite{tis2} ($\mathcal{SG}\mathbf{227}$) is  a glide-protected TCI with hourglass surface states. Second, elemental phosphorus in the A7 structure ($\mathcal{SG}\mathbf{166}$), which occurs at about 9GPa \cite{P}, is predicted to be an inversion-protected TCI with 1D hinge states, akin to the recently realized case of elemental bismuth \cite{Bi}.
Finally, we find that Ag$_{2}$F$_{5}$ \cite{ag2f5} ($\mathcal{SG}\mathbf{2}$) is a weak TI with   additional inversion-protected band topology characterized by the invariant $\kappa_1=2$.

\textit{Centrosymmetric strong TIs.}
Our search algorithm also naturally identifies conventional strong TIs diagnosable through the Fu-Kane parity criterion \cite{Fu-Kane}.
We identify
CaAs$_{3}$ ($\mathcal{SG}\mathbf{2}$), Bi$_{2}$PbTe$_{4}$ ($\mathcal{SG}\mathbf{166}$) and  CaGa$_2$As$_2$ ($\mathcal{SG}\mathbf{166}$) are all
STIs \cite{SM}.

\textit{Other types of topological semimetals.}
We also found other topological semimetals corresponding to case 3 described in Sec.\ \ref{sec:SI-theory}. In particular, AuLiMgSn and AgF$_{2}$ are particularly interesting, with the former featuring symmetry-enforced band crossings leading to three-fold degenerate gaplessness points, albeit at $\sim 0.3$ eV above the Fermi energy, and the latter realizing a nodal line semimetal. The electronic structures and symmetry analysis are presented in the SM \cite{SM}.

\section{Conclusion}
We proposed an efficient and versatile method to search for topological (crystalline) materials
by combining the theory of symmetry indicators of band topology combined with first
principles calculations.
Focusing on merely $8$  space groups, we discovered numerous topological materials featuring a diverse set of band topology, which ranges from conventional $\mathbb Z_2$ topological insulators and nodal semimetals to recently proposed phases of topological crystalline insulators.
Many more symmetry settings are yet to be explored. To wit, there are 230 space groups, and even $1,651$ magnetic space groups, for which the symmetry indicators have been completely derived \cite{Po, Po-2}. It is of great interest to apply our materials search strategy to these other symmetry settings, where many more topological materials likely reside.\\

\noindent{\it Note Added:} While completing this manuscript, Ref.\ \onlinecite{Ca2As} appeared, which proposed a candidate material for a rotation-symmetry protected TCI  confirmed using symmetry indicators.

\begin{acknowledgements}
HCP and AV thank E.\ Khalaf and H.\ Watanabe for earlier collaborations on related topics.
FT and XGW were supported by National Key R\&D Program of China (No.\ 2017YFA0303203 and 2018YFA0305700), the NSFC (No.\ 11525417, 51721001 and 11790311). FT was also
supported by the program B for Outstanding PhD candidate of Nanjing
University.
XGW was partially supported by a QuantEmX award funded by the Gordon and Betty Moore Foundation's EPIQS Initiative through ICAM-I2CAM, Grant GBMF5305 and by the  Institute of Complex Adaptive Matter (ICAM).
AV is supported by NSF DMR-1411343, a Simons Investigator Grant, and by the ARO MURI on topological insulators, grant W911NF- 12-1-0961.
\end{acknowledgements}

\bibliography{refs}

\end{document}



\clearpage
\setcounter{section}{0}
\onecolumngrid
\begin{center}
{\large \bf Supplementary Materials}
\end{center}
\tableofcontents
\section{Detailed descriptions on implementation of SI method.}

Symmetry indicators (SIs) \cite{Po} of band topology are very powerful for
the efficient diagnosis and prediction of topological materials based on
the first principles calculations. In this section, we will give  detailed
descriptions  on how to implement the SI method to a real material, and
diagnose it by first principles calculations. Other than an indicator of
band topology for the insulator, SI method is also very powerful to diagnose
the topological semimetal (e.g.  Dirac semimetal (DSM), multiple-fold
degenerate semimetal, nodal-line semimetal, etc.) as we will show in the following section.

One first performs a routine first principles electronic structure calculation for a
material. Then according to the space group ($\mathcal{SG}$), one can obtain
the irreducible representations (irreps) of its  valence bands (i.e. the first $\nu$ bands, where $\nu$ is the number of valence electrons in the primitive unit cell) at all
the high symmetry points (HSPs) in the Brillouin zone (BZ). We denote the little group at the $i$th  HSP, namely $k_{i}$ point, as $\mathcal{G}(\mathbf{k}_{i})$. The $\alpha_i$th ($\alpha_i=1,2,\ldots$) irreducible representation (irrep) of
$\mathcal{G}(\mathbf{k}_{i})$ is labeled by $\alpha_{i}$. If the
valence bands and conduction bands are separated throughout the BZ , the
electronic structure can be described by the number of the occurrences for
the $\alpha_{i}$th irrep in the valence  electronic bands, i.e. $n_{\mathbf{k}%
_{i}}^{\alpha _{i}}$. The symmetry content of such valence bands is dubbed a ``band
structure'' (BS) in Ref.\ \onlinecite{Po}. The BS can be represented  by an integer-valued vector, $\mathrm{%
\mathbf{n}}=$:
\begin{eqnarray}
&&(\nu,n_{\mathbf{k}_{1}}^{1},n_{\mathbf{k_{1}}%
}^{2},\ldots ,n_{\mathbf{k}_{\mathbf{1}}}^{\alpha _{1}},\ldots ,n_{\mathbf{k}%
_{1}}^{r_{1}},n_{\mathbf{k}_{2}}^{1},n_{\mathbf{k_{2}}}^{2},\ldots ,n_{%
\mathbf{k}_{2}}^{\alpha _{2}}\ldots ,n_{\mathbf{k}_{2}}^{r_{2}},
\label{app:n} \\
&&n_{\mathbf{k_{i}}}^{1},\ldots ,n_{\mathbf{k_{i}}}^{\alpha _{i}},\ldots ,n_{%
\mathbf{k_{i}}}^{r_{i}},\cdots ,n_{\mathbf{k}_{N}}^{1},\ldots ,n_{\mathbf{%
k}_{N}}^{\alpha_{N}},\ldots ,n_{\mathbf{k}_{N}}^{r_{N}}).
\end{eqnarray}%
In the above equation, $\mathbf{k}_{i}$ denotes the HSP as before, where $i$ takes $%
1,2,\ldots ,N$ ($N$ is the total number of HSPs). The superscript
labels the irrep for the corresponding $\mathcal{G}(\mathbf{k}_{i})$, $\alpha_i=1,2,\ldots,r_i$ and the
number of these irreps is $r_{i}$ for the $i$th HSP. All the HSPs and their
irreps for the 230 $\mathcal{SG}s$ can be found in Ref. \cite{Bradley}.

Suppose that the valence and conduction bands touch at the $i$th HSP (i.e. $\mathbf{k}%
_{i}$), and these touching  bands form the $j$th irrep of $\mathcal{G}(\mathbf{k}_{i})$.
We can still obtain $n_{\mathbf{k}_{i}}^{j}$ by the standard method. In this case $n_{\mathbf{k}_i}^j$ is not generally an integer. This  belongs to  Case 3 in the main text. Furthermore, even if all the ${n_{\mathbf{k}}^{\alpha }}$'s are integers, they may not satisfy all the compatibility relations, i.e., there will be some symmetry-enforced
band crossing(s) in high symmetry line or plane. This also belongs to Case 3 of the main text. When there is an indicator of the band crossing, we then need to carefully analyze  the position(s) of the band crossing(s).

On the other side, it is clear that the atomic insulator (AI), namely a group of electronic  bands  adiabatically connected to a strict atomic limit, forms a BS. Ref.\ \onlinecite{Po} has shown that there are $d_{\mathrm{AI}}^{\mathcal{SG}}$  AI basis vectors (i.e. $\mathbf{a}_{i}^{\mathcal{SG}}$)  for any given $\mathcal{SG}$, i.e. any
AI in this $\mathcal{SG}$ can be\ expressed by a linear combination of $%
d_{\mathrm{AI}}^{\mathcal{SG}}$ AI basis vectors \cite%
{Po}: $\mathrm{\mathbf{n}}^\mathrm{AI}=\sum_{i=1}^{d_{\mathrm{AI}}^{^{%
\mathcal{SG}}}}m_{i}\mathbf{a}_{i}^{^{\mathcal{SG}}}$, where $m_{i}'s$ are all
integers \cite{Po} as Eq. 2 of the main text. (In Sec. \ref{AI} , we give the detailed descriptions on
how to calculate the AI basis vectors for a given $\mathcal{SG}$. We also list the
AI basis vectors for all the $\mathcal{SG}s$ encountered in the work in Sec. \ref{AIS}).

In addition to  AI,  Ref.\ \onlinecite{Po} also found that any BS in $\mathcal{SG}$
can also be expanded on the AI basis vectors:
\begin{equation}
\mathrm{\mathbf{n}}=\sum_{i=1}^{d_{\mathrm{AI}}^{\mathcal{SG}}}q_{i}\mathbf{a%
}_{i}^{\mathcal{SG}},  \label{app:bsonai}
\end{equation}
however here the expansion coefficients $q_i's$ can be non-integers as we show in the main text. This is because that some AI basis vector $\mathbf{a}_i$ may have a common factor $C_i$ for all its entries, so $q_i$ can be a rational number only requiring that $q_iC_i$ is an integer. When all the ${q_i}'s$ are integers, the BS is indistinguishable from an AI as far as symmetry representations are concerned (in the ``stable sense'' as elaborated in Ref.\ \onlinecite{Fragile}). It should be emphasized that this alone does not preclude the existence of band topology; rather, it simply implies more refined methods are required to detect, or rule out, band topology in the system.

In contrast, when some $q_i$'s are not integers but $q_iC_i$'s are all integers, up to detailed energetics the system is a topological (crystalline) insulator (here, as in the main text, we consider systems with time-reversal symmetry and strong spin-orbit coupling; the statements have to be modified accordingly in other symmetry settings \cite{Po-2,Fang-4}).

The AI basis vectors for any $\mathcal{SG}$ can be easily obtained, and the first principles calculations for $n_{\mathbf{k}}^\alpha$ is computationally easy as it only involves wave-function data at a small number of isolated momenta.  Thus, simply by analyzing the expansion coefficients $q_i's$, it is  highly efficient to screen crystal-structure databases and  diagnose the topological (crystalline) insulators or (semi-)metals, following the flow described in Fig. 1 in the main text.

\section{The detailed discussions on materials highlighted in the main text}

\begin{figure}[tbp]
\includegraphics[width=19cm]{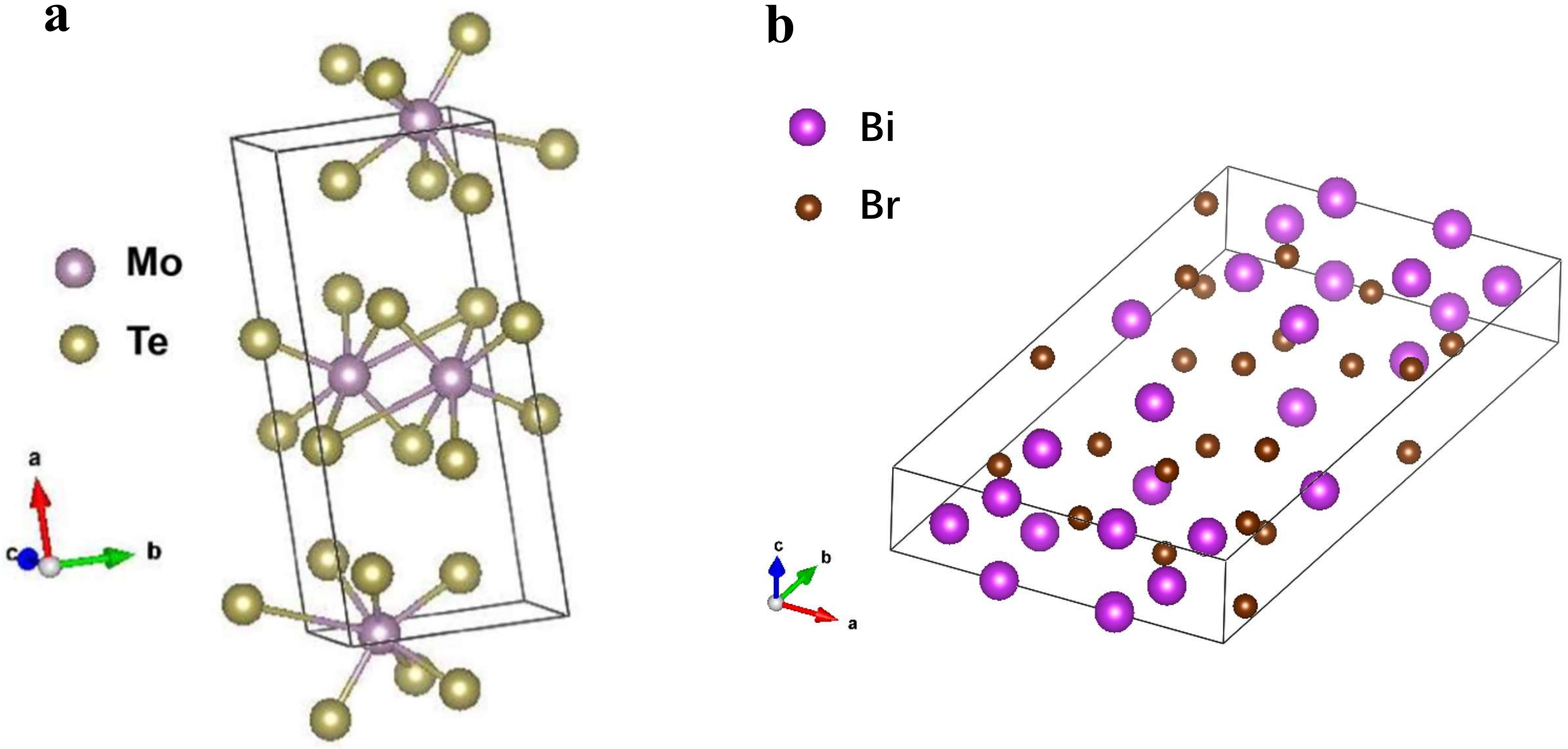}\newline
\caption{The crystal structures for $\protect\beta$-MoTe$_2$ (a) and BiBr
(b). }
\label{app:stru}
\end{figure}

\subsection{$\protect\beta$-MoTe$_2$}

The $\beta $-MoTe$_{2}$ \cite{mote2-beta} crystallizes in the primitive monoclinic Bravais lattice. Its $\mathcal{SG}$ is No. $11$. The crystal structure is shown in Fig. \ref{app:stru}(a). We
adopt such a setting that  $\mathbf{c}$ is the unique axis, i.e. the $C_{2}$ screw rotation axis is along $\mathbf{c}$. We use the experimental structural data for the structural parameters \cite{mote2-beta}. There are 2 inequivalent Mo's and 4 inequivalent Te's, and
they all occupy $2e$ Wyckoff positions. Hence the multiplicity of the
chemical formula units is 4, and there are in total 12 atoms in the
primitive cell. There are in total 56 valence electrons in the primitive unit cell, i.e. $\nu=56$. We list the HSPs, the irreps for each HSP, and the first principles calculated numbers $n_{\mathbf{k}_i}^{\alpha_i}$ in Table \ref{mote2-n}:
\begin{table}[!htbp]
\centering
\caption{For $\mathcal{SG}\mathbf{11}$, the HSPs are given by the labels $\Gamma,B,\ldots$ in order. For the labeling of the irreps of $\mathcal{G}(\mathbf{k}_i)$, we use $(j,m)$ where $j$ means the $j$th irrep and $m$ denotes the dimension of the corresponding irrep. They are all listed in  Ref. \cite{Bradley}. We use the same order of the irrep as Ref. \cite{Bradley}. The red color means that due to $\mathcal{T}$, the irrep must occur with its $\mathcal{T}$ pair (belonging to the same irrep) simultaneously. Thus $\mathcal{T}$ requires that the red colored irreps must happen even times. So it is necessary to divide them by 2 \cite{Po} to obtain the physical common factors.}
\begin{tabular}{|c|c|c|c|c|c|c|c|c|c|c|c|c|c|c|c|c|c|c|c|c|}
  \hline
   HSP& \multicolumn{4}{|c|}{$\Gamma$}& \multicolumn{4}{|c|}{$B$}&\multicolumn{4}{|c|}{$Y$}&$Z$&$C$&$D$&\multicolumn{4}{|c|}{$A$}&$E$\\ \hline
   irrep&(1,1)&(2,1)&(3,1)&(4,1)&(1,1)&(2,1)&(3,1)&(4,1)&(1,1)&(2,1)&(3,1)&(4,1)&{\color{red}{(1,2)}}&{\color{red}{(1,2)}}&{\color{red}{(1,2)}}&(1,1)&(2,1)&(3,1)&(4,1)&
   {\color{red}{(1,2)}}\\ \hline
   $n_{\mathbf{k}_i}^{\alpha_i}$& 16&16&12&12&14&14&14&14&14&14&14&14&14&14&14&14&14&14&14&14\\ \hline
\end{tabular}
\label{mote2-n}
\end{table}
According to  Table \ref{mote2-n}, we see that for HSPs $\Gamma,B,Y,A$, they all have four 1D irreps while the rest HSPs have one 2D irrep. Thus there are in total 20 ${n_{\mathbf{k}^\alpha}'s}$. Consider the filling number $\nu$ in the considered bands \cite{Po}, the total number of the entries for any BS in $\mathcal{SG}\mathbf{11}$ is 21.     Form Table \ref{mote2-n}, one can also readily find that the valence bands of $\beta$-MoTe$_2$ constitute the BS as follows:
\begin{eqnarray}
&&\mathrm{\mathbf{n}}%
=(56,16,16,12,12,14,14,14,14,14,14,14,14,14,14,14,14,14,14,14,14)
\label{app:mote-bs} \\
&=&12\mathbf{a}_{1}+2\mathbf{a}_{2}+\mathbf{a}_{3}+\mathbf{a}_{4}+\frac{1}{2}%
\mathbf{a}_{5},
\end{eqnarray}%
where $\mathbf{a}_{5}$ has common factor 4, $\mathbf{a}_{3,4}$ have common
factor 2 while the others have no common factor as shown in Sec. \ref{AIS}.  The expansion on the AI basis vectors is $%
q=(12,2,1,1,\frac{1}{2})$. The SI for $\beta$-MoTe$_2$ is thus (0,0,2), and this indicates a nontrivial topology of band \cite{Po}. \\

Based on the first principle calculations, we
also obtain the parities of  the valence bands, and list the number of even($+$)/odd($-
$) parity in Table \ref{app:parity-mote2}.
\begin{table}[htbp!]
\centering
\begin{tabular}{|c|c|c|c|c|c|c|c|c|}
\hline
$\mathbf{k}\in$ TRIM & $\Gamma$ & $X$ & $Y$ & $Z$ & $U$ & $T$ & $S$ & $R$ \\
\hline
$n_{\mathbf{k}}^+$ & 16 & 14 & 14 & 14 & 14 & 14 & 14 & 14 \\ \hline
$n_{\mathbf{k}}^-$ & 12 & 14 & 14 & 14 & 14 & 14 & 14 & 14 \\ \hline
\end{tabular}\label{app:parity-mote2}
\caption{The calculated parities for the valence bands of $\protect\beta $%
-MoTe$_{2}$. $n_{\mathbf{k}}^{\pm }$ is the number of the occupied even/odd
Kramers pairs (KPs), respectively.}
\end{table}
Based on the obtained band parities, one can find that the conventional 3D
topological indices, i.e. $(\nu _{0};\nu _{1},\nu _{2},\nu _{3})$ \cite{Fu-Kane}, are all vanishing. However,  the newly-introduced inversion
topological invariant $\kappa_{1}$ \cite{Khalaf,Fang-3} is nonvanishing. It is defined by \cite{Khalaf,Fang-3}:
\begin{equation}
\kappa _{1}=\sum_{\mathbf{k}\in \mathrm{TRIM}}(n_{\mathbf{k}}^{+}-n_{\mathbf{%
k}}^{-})/2\quad \mathrm{mod}\quad 4.  \label{app:kappa1}
\end{equation}%
It is easy to find that  $\kappa_1=2$ according to Table \ref{app:parity-mote2}. Based on Eq. (\ref{app:kappa1}) and the Fu-Kane formula  for $(\nu
_{0};\nu _{1},\nu _{2},\nu _{3})$ \cite{Fu-Kane}, one can easily understand
the possible topological phase transition in $\beta $-MoTe$_{2}$ and the phase diagram
illustrated in Fig. 2(a) of the main text. Through adjusting the lattice structure by strain/pressure, one can tune the parities of the occupied bands. For example,  a parity switch occurs at $\Gamma$ point that the occupied bands are changed to own 15 even KPs and 13 odd KPs. A topological transition to a strong topological insulator phase occurs. By this way, $\beta$-MoTe$_2$ can further be tuned to be a trivial insulator by an additional parity switch. Thus for $\beta$-MoTe$_2$ the possible topological phase transition induced by strain/pressure is highly interesting.   \newline
\begin{figure}
  \includegraphics[width=19cm]{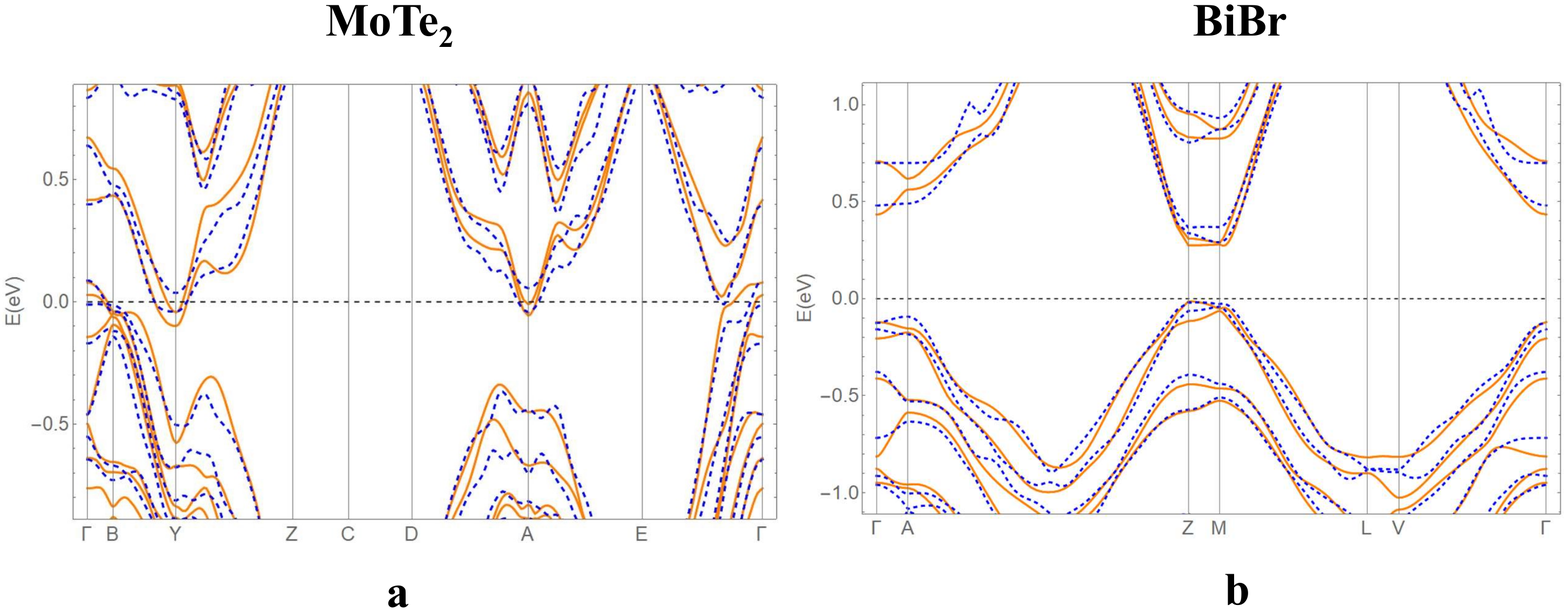}\\
  \caption{The comparison between the first principles calculated electronic bands (solid line) and the TB bands (dashed line) for (a) $\beta$-MoTe$_2$ and (b) BiBr.}\label{app:fit}
\end{figure}
We construct a tight-binding (TB) model for $\beta$-MoTe$_2$ in Sec. \ref{TB}. This TB model not only reproduces the energy bands reasonably shown in Fig. \ref{app:fit}(a), it also gives exactly the same ${n_{\mathbf{k}}^\alpha}'s$ and mirror Chern numbers calculated by first principles calculations. Based on the TB model we demonstrate the hinge states by constructing a centrosymmetric  prism along $\mathbf{c}$, and calculating the prism's electronic structure.  To unambiguously distinguish the hinge states from the bulk and surface states, we plotted  the real space distributions of the eigen-states of the prism. In Fig. \ref{app:hinge}, we show the real space distribution of  two of the four hinge modes at the Fermi level (The other two are related to them by $\mathcal{T}$). These hinge states have their directions of spin locked to their motions forming a helical pattern as shown in Fig. 2(c) in the main text.
\begin{figure}[htbp!]
\includegraphics[width=20cm]{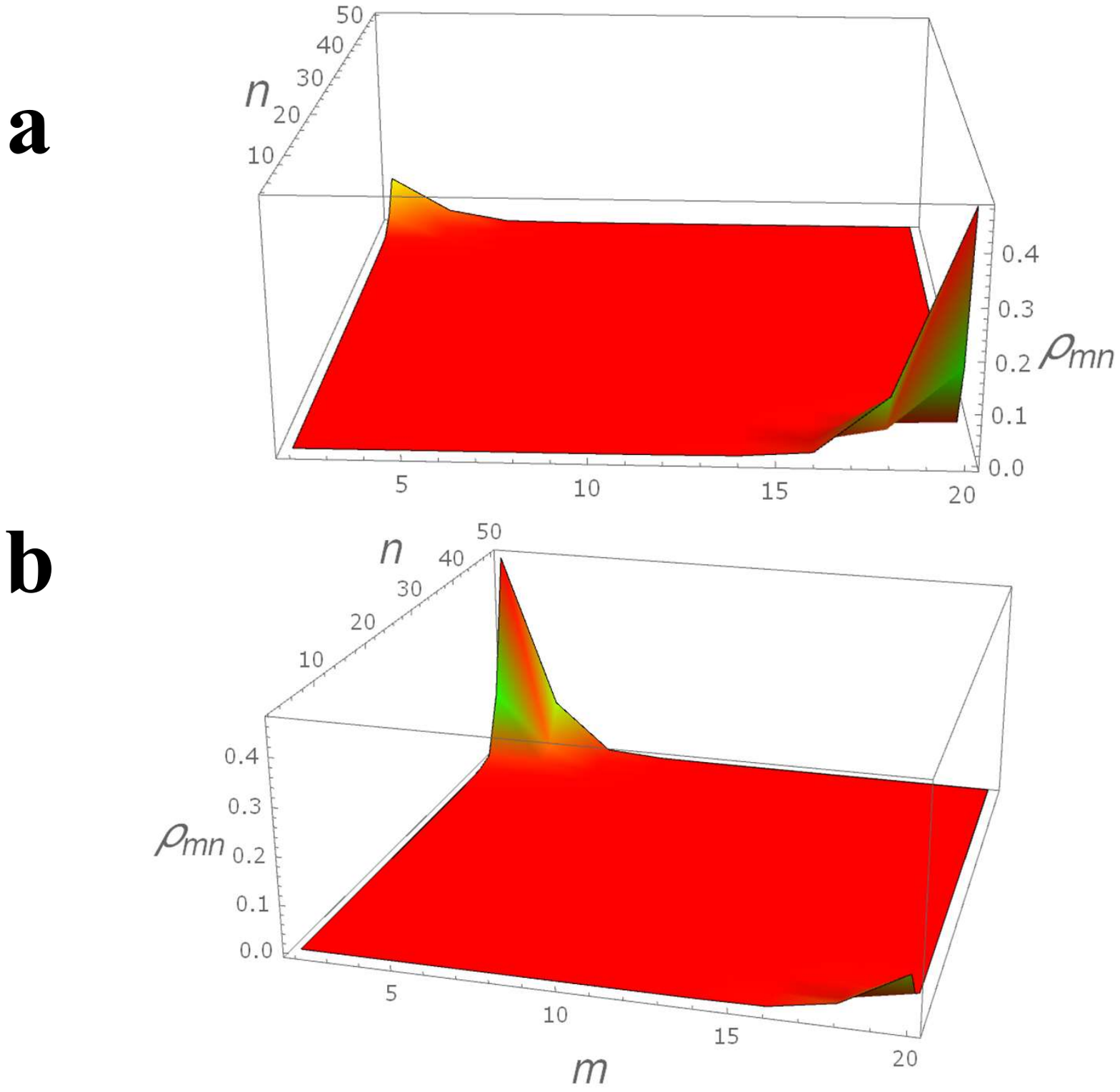}\newline
\caption{The real space distribution for the two hinge states $\psi^{1},\psi^{2}$  at the Fermi level (with positive group velocities, See red lines in Fig. 2(d) of the main text for their dispersions) and the other two hinge states are related to them by $\mathcal{T}$ . $m,n$ label the position while $\protect\rho_{mn}\equiv\sum_{\protect%
\mu}|{\protect\psi^{1,2}(m,n,\protect\mu)}|^2$ where
summation is over $\protect\mu$, the collected set of the sublattice,
orbital and spin. }
\label{app:hinge}
\end{figure}

\subsection{BiBr}

\begin{figure}[htbp!]
\includegraphics[width=20cm]{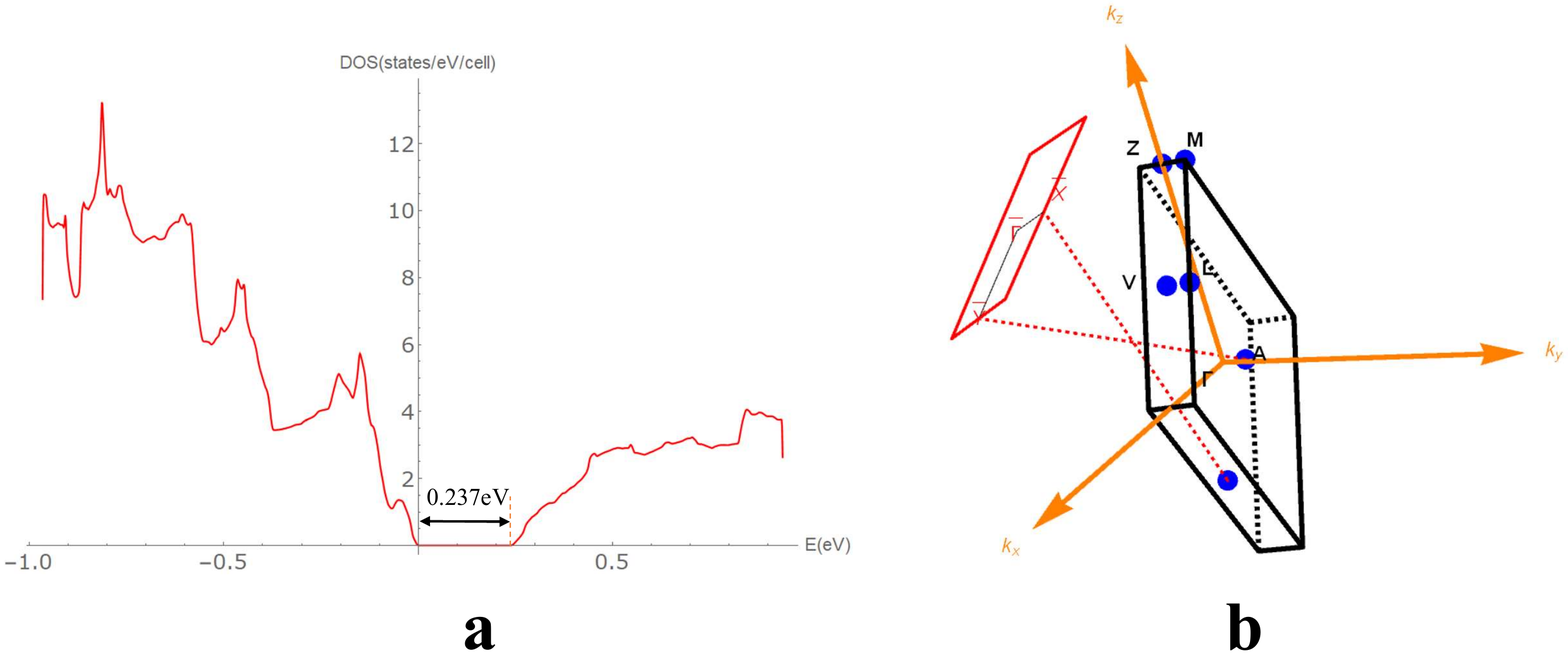}\newline
\caption{(a) The DOS plot for BiBr; (b) The BZ of BiBr and the (001)-surface BZ with the projection of the surface $\bar{X}$ and $\bar{Y}$ k point from the bulk momenta indicated by the dashed red line: note that we choose the primitive unit cell for the BZ rather than the Wigner-Seitz cell.}
\label{app:SM-dos+bz}
\end{figure}
$\mathcal{SG}\mathbf{11}$ has a 2-fold screw axis which cannot protect any surface states \cite{FangFu} while for the neighbor $\mathcal{SG}\mathbf{12}$, it has symmorphic $C_2$ rotation. This symmetry can protect surface Dirac cones in $C_2$ symmetric surfaces \cite{FangFu}. We search $\mathcal{SG}\mathbf{12}$, and  we found another \textquotedblleft 2 in $\mathbb{Z}_{4}$%
\textquotedblright\ material: BiBr  \cite{bibr} with $\mathcal{SG}\mathbf{12}$. We choose  $\mathbf{c}$ as the unique axis (i.e. $\mathbf{c}$ is the $C_2$ rotation axis). The fundamental
lattice basis vectors can then be chosen as,
\begin{eqnarray}
&&\mathbf{a}_{1}=\frac{1}{2}(\mathbf{a}-\mathbf{c}),  \label{app:sg12-a1a2a3}
\\
&&\mathbf{a}_{2}=\mathbf{b}, \\
&&\mathbf{a}_{3}=\frac{1}{2}(\mathbf{a}+\mathbf{c}), \\
&&
\end{eqnarray}%
where $\mathbf{a,b,c}$ are conventional unit cell vectors.
There are four inequivalent Bi's and four inequivalent Br's, and they all
occupy the $4i$ Wyckoff position. So the total number of atoms in the primitive unit cell is 16 (note that the Bravais is of the base-centered type). The calculated BS for the 64 valence
bands are given by,
\begin{table}
\centering
\caption{For $\mathcal{SG}\mathbf{12}$, the HSPs are given by the labels $\Gamma,A,\ldots$ in order, and their coordinates can be referred to Ref. \cite{Bradley}. For the labeling of the irreps of $\mathcal{G}(\mathbf{k}_i)$, we use $(j,m)$ where $j$ means the $j$th irrep as listed in order by Ref. \cite{Bradley} and $m$ denotes the dimension. The red color means that due to $\mathcal{T}$, the irrep must occur simultaneously with its $\mathcal{T}$ pair which belongs to the same irrep.}
\begin{tabular}{|c|c|c|c|c|c|c|c|c|c|c|c|c|c|c|c|c|c|c|c|c|}
  \hline
   HSP& \multicolumn{4}{|c|}{$\Gamma$}& \multicolumn{4}{|c|}{$A$}&\multicolumn{4}{|c|}{$Z$}&\multicolumn{4}{|c|}{$M$}&\multicolumn{2}{|c|}{$L$}&\multicolumn{2}{|c|}{$V$} \\ \hline
   irrep&(1,1)&(2,1)&(3,1)&(4,1)&(1,1)&(2,1)&(3,1)&(4,1)&(1,1)&(2,1)&(3,1)&(4,1)&(1,1)&(2,1)&(3,1)&(4,1)&{\color{red}{(1,1)}}&{\color{red}{(2,1)}}&
   {\color{red}{(1,1)}}&{\color{red}{(2,1)}}\\ \hline
   $n_{\mathbf{k}_i}^{\alpha_i}$& 18&18&14&14&16&16&16&16&16&16&16&16&16&16&16&16&16&16&16&16\\ \hline
\end{tabular}
\end{table}
\begin{eqnarray}
&&\mathrm{\mathbf{n}}%
=(64,18,18,14,14,16,16,16,16,16,16,16,16,16,16,16,16,16,16,16,16)
\label{app:bibir-bs} \\
&=&14\mathbf{a}_{1}+2\mathbf{a}_{2}+2\mathbf{a}_{3}+2\mathbf{a}_{4}+\mathbf{a%
}_{6}-\frac{1}{2}\mathbf{a}_{7},
\end{eqnarray}%
where $\mathbf{a}_{7}$ has common factor 4, $\mathbf{a}_{5,6}$ have common
factor 2 while the others have no common factor as shown in Sec. \ref{AIS}. The  density of states (DOS) is plotted in Fig. \ref{app:SM-dos+bz}, which
 shows that there is a full gap ($\sim 237$meV).

 The SI is (0,0,2) for BiBr, and in this case $C_{2}$
protects surface Dirac cones. We construct a TB model as shown in Sec. \ref{TB} with its electronic energy spectrum shown in Fig. \ref{app:fit}(b). It also reproduces the same BS and mirror Chern numbers. Based on this model, we construct a slab and solve for the surface states. The slab geometry is finite
 along $\mathbf{a}_{3}$ with its infinite plane parallel to
(001) (according to the conventional unit cell) plane. For this slab, the
supercell unit vectors can be chosen as $\mathbf{\bar{a}_{1}}=\mathbf{a}$
and $\mathbf{\bar{a}_{2}}=\mathbf{b}$. It is a oblique cell, whose
convenient BZ can be chosen as the primitive one rather than the
Wigner-Seitz cell: the oblique tetragon spanned by, $\mathbf{\bar{b}_{1}},%
\mathbf{\bar{b}_{2}}$:
\begin{eqnarray}
&&\mathbf{\bar{b}_{1}}=2\pi \frac{\mathbf{\bar{a}_{2}}\times \mathbf{a_{3}}}{%
\mathbf{a_{3}}\cdot (\mathbf{\bar{a_{1}}}\times \mathbf{\bar{a_{2}}})},
\label{app:surfacebz} \\
&&\mathbf{\bar{b}_{2}}=-2\pi \frac{\mathbf{\bar{a}_{1}}\times \mathbf{a_{3}}%
}{\mathbf{a_{3}}\cdot (\mathbf{\bar{a_{1}}}\times \mathbf{\bar{a_{2}}})}.
\end{eqnarray}%

For each $\mathbf{\bar{k}}$ for this surface BZ, $\mathbf{\bar{k}}=k_{1}%
\mathbf{\bar{b}_{1}}+k_{2}\mathbf{\bar{b}_{2}}$. The surface BZ is plotted
in Fig. \ref{app:SM-dos+bz}, with the projection from the bulk BZ indicated.\newline
For the calculation of the hinge states, we constructed a  prism along $%
\mathbf{c}$. The other two sides of this prism are chosen along $\mathbf{b}$
and $\mathbf{a}_{3}$, and they are both open.  The same as  $\beta$-MoTe$_2$, based on the TB model, we solve for the eigen-solutions for this prims.  We distinguish the hinge states through the real space distributions of the eigen-states of the prism.

\subsection{AgXO,X=Na,K,Rb}

\begin{figure}
  \includegraphics[width=20cm]{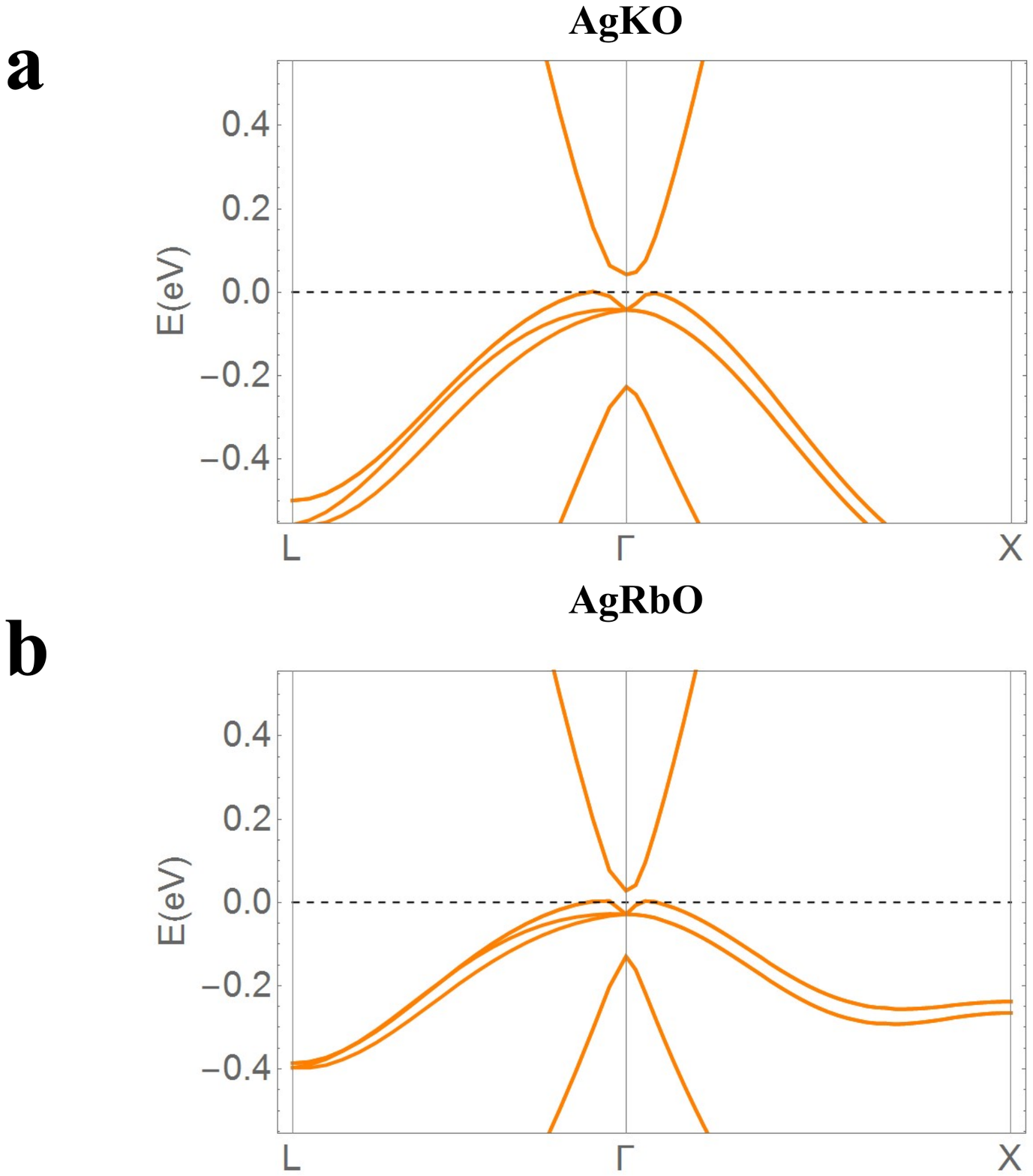}\\
  \caption{The electronic band plots for two noncentrosymmetric strong topological insulators: (a) AgKO and (b) AgRbO.}\label{app:fig-AgXO}
\end{figure}

Other than the centrosymmetric $\mathcal{SG}s$, we also search those
noncentrosymmetric ones but with $\mathcal{S}_4$ symmetry. Focusing on $\mathcal{SG}\mathbf{216}$, we found that a series of materials, AgXO with X=Na,K,Rb,
are  1 in $\mathbb{Z}_2$ and the electronic band plots for X=K and
Rb are shown in Fig. \ref{app:fig-AgXO}.  1 in
such $\mathbb{Z}_2$ indicates nontrivial topology and further, it corresponds to a 3D STI \cite{Khalaf,Fang-3}. For the centrosymmetric systems, one can use the
Fu-Kane criteria \cite{Fu-Kane} to judge a 3D topological phase. A similar
criteria actually exists for the $S_4$ symmetric systems \cite{Khalaf,Fang-3}. The corresponding topological invariant is defined by \cite{Khalaf,Fang-3}:
\begin{equation}  \label{app:kappa4}
\kappa_4=\frac{1}{2\sqrt2}\sum_{\mathbf{k}\in \mathrm{SIM},s_4}s_4n_\mathbf{k%
}^{s_4}\quad\mathrm{mod}\quad2,
\end{equation}
where SIM represents the set of $S_4$ invariant momenta and $s_4$ is the
eigenvalue of $S_4$: $s_4=e^{-i\frac{j_z\pi}{2}},j=\pm \frac{1}{2},\pm \frac{3}{2}$. Due to $\mathcal{T}$
symmetry, $s_4$ must occurs at the same time with $s_4^*$, though $s_4^*$
maybe at a different $\mathbf{k}$ in the SIM. The $\mathcal{SG}\mathbf{216}$ here has a face-centered cubic lattice, whose four SIM are $\Gamma,Z,P',P$. They are listed in the Table \ref{app:s4}, where  we also give the number of the occurrences for each  eigenvalue $s_4$ for the 18 valence bands at SIM by first principles calculations. Note that the band inversion happens near $\Gamma$ point, and when $n_{\Gamma}^{\pm\frac{1}{2}}$ is changed to be 4 (and $n_{\Gamma}^{\pm \frac{3}{2}}$ would be 5), $\kappa_4$ becomes zero and the materials become a trivial insulator.

\begin{table}[htbp!]
\centering
\begin{tabular}{|c|c|c|c|c|}
\hline
$\mathbf{k}\in$ SIM & $\Gamma$ & $Z$ & $P'$ & $P$ \\
\hline
$n_{\mathbf{k}}^{\frac{1}{2}}$ &  5& 4& 4& 4\\ \hline
$n_{\mathbf{k}}^{-\frac{1}{2}}$ &  5& 4& 4& 4\\ \hline
$n_{\mathbf{k}}^{\frac{3}{2}}$ &   4& 5& 5& 5\\ \hline
$n_{\mathbf{k}}^{-\frac{3}{2}}$ &  4& 5& 5& 5\\ \hline
\end{tabular}%
\caption{The calculated parities for the valence bands of AgXO. $n_{\mathbf{k}}^{j_z} $ is the number of the occupied eigen-states with $S_4$ eigenvalue $e^{-i\frac{j_z\pi}{2}}$. The coordinates of the four SIM are $\Gamma=(0,0,0),Z=(0,0,1),P'=(0,1,-\frac{1}{2}),P=(0,1,\frac{1}{2})$ respectively  in the conventional and reciprocal basis vectors.}
\label{app:s4}
\end{table}

\subsection{$\mathbf{k}\cdot\mathbf{p}$ model for the Dirac
semimetal in the main text}
For the Dirac semimetal Bi$_2$MgO$_6$, we derive the low energy $\mathbf{k}%
\cdot\mathbf{p}$ Hamiltonian as follows: Consider the Hamiltonian $H(\mathbf{%
k})$ in $\Gamma$-$Z$ line, it is subject to the symmetry restriction of $C_{4z}$ and
$\sigma_v$. We choose $\sigma_v$ is perpendicular to $y$ axis.  The band inversion occurs in  $\Gamma$-$Z$
between two bands near the Fermi level shown in Fig. 4 of the main text. The
two bands belong to $\Lambda_6$ and $\Lambda_7$ irreps respectively of the
symmetry group $C_{4v}$. For the $\Lambda_6$ band,  the two basis
vectors are $|j_z>$ ($j_z=\pm \frac{1}{2}$) while for the $\Lambda_7$ band,
the two basis vectors are $|j_z>$ with $j_z=\pm \frac{3}{2}$. $j_z$ means
that the eigenvalue of $C_{4z}$ is $e^{-i\frac{j_z\pi}{2}}$. $\sigma_v$
relate $|j_z>$ to $|-j_z>$. So each band is two-fold degenerate, which is also dedicated by the joint operation $\mathcal{TI}$. Hence the representation matrix for $C_{4z}$ is $\mathrm{dia}(e^{-i\frac{\pi}{4}},e^{i\frac{\pi}{4}},e^{-i\frac{3\pi}{4}},e^{i%
\frac{3\pi}{4}})$ and for $\sigma_v$, it is $i\sigma_0\otimes\sigma_1$ (note $\sigma_v^2=-1$). In
this line and under the $\Lambda_6$ and $\Lambda_7$ band representations, $H(%
\mathbf{k})=M(\mathbf{k})\sigma_3\otimes\sigma_0$. So in $\Gamma$-$Z$, when $M(\mathbf{k})=0$ has a solution $\mathbf{k}=\mathbf{k}_D$
, a band crossing happens at this point.  The low energy Hamiltonian to the linear term around this point is written
as $h(\mathbf{q})=\frac{\partial{H(\mathbf{k}_D)}}{\partial \mathbf{k}_D}
\cdot \mathbf{q}\equiv \mathbf{H}^{(1)}\mathbf{q}$. $h(\mathbf{q})$ is
restricted by $C_{4z}$ and $\sigma_v$ that
\begin{eqnarray}  \label{app:dsm-kp}
&& \mathrm{D}(C_{4z})^\dag \mathbf{H}^{(1)} \mathrm{D}(C_{4z})=C_{4z}\mathbf{%
H}^{(1)}, \\
&& \mathrm{D}(\sigma_v)^\dag \mathbf{H}^{(1)} \mathrm{D}(\sigma_v)=\sigma_v%
\mathbf{H}^{(1)}.
\end{eqnarray}
$\mathcal{T}\mathcal{I}=UK$ also has effect to the $\mathbf{k}\cdot\mathbf{p}
$ Hamiltonian by $(U^\dag \mathbf{H}^{(1)}U)^*=\mathbf{H}^{(1)}$ where $%
U=\sigma_0\otimes \sigma_2$. We finally obtain the symmetry allowed $\mathbf{%
k}\cdot\mathbf{p}$ Hamiltonian as follows (in the basis of $\{|\frac{1}{2}>,|\frac{3}{2}>,|-\frac{1}{2}>,|-\frac{3}{2}>\}$):

\begin{equation}\label{app:dkp}
    h(\mathbf{q})=\left(\begin{array}{cc}
    c_0 q_3\sigma_0+c'q_3\sigma_3+c(q_1\sigma_1+q_2\sigma_2)& 0\\
    0&   c_0 q_3\sigma_0+c'q_3\sigma_3+c(q_1\sigma_1-q_2\sigma_2)
    \end{array}\right),
\end{equation}
where the two $2\times2$ block matrices correspond to two Weyl equations of opposite chirality regardless of  a $q_z$ dependent constant term.

\section{The other ``2 in $\mathbb{Z}_4$'' materials}

\subsection{Hourglass insulator TiS$_2$}

\begin{figure}[!htbp]
\includegraphics[width=15cm]{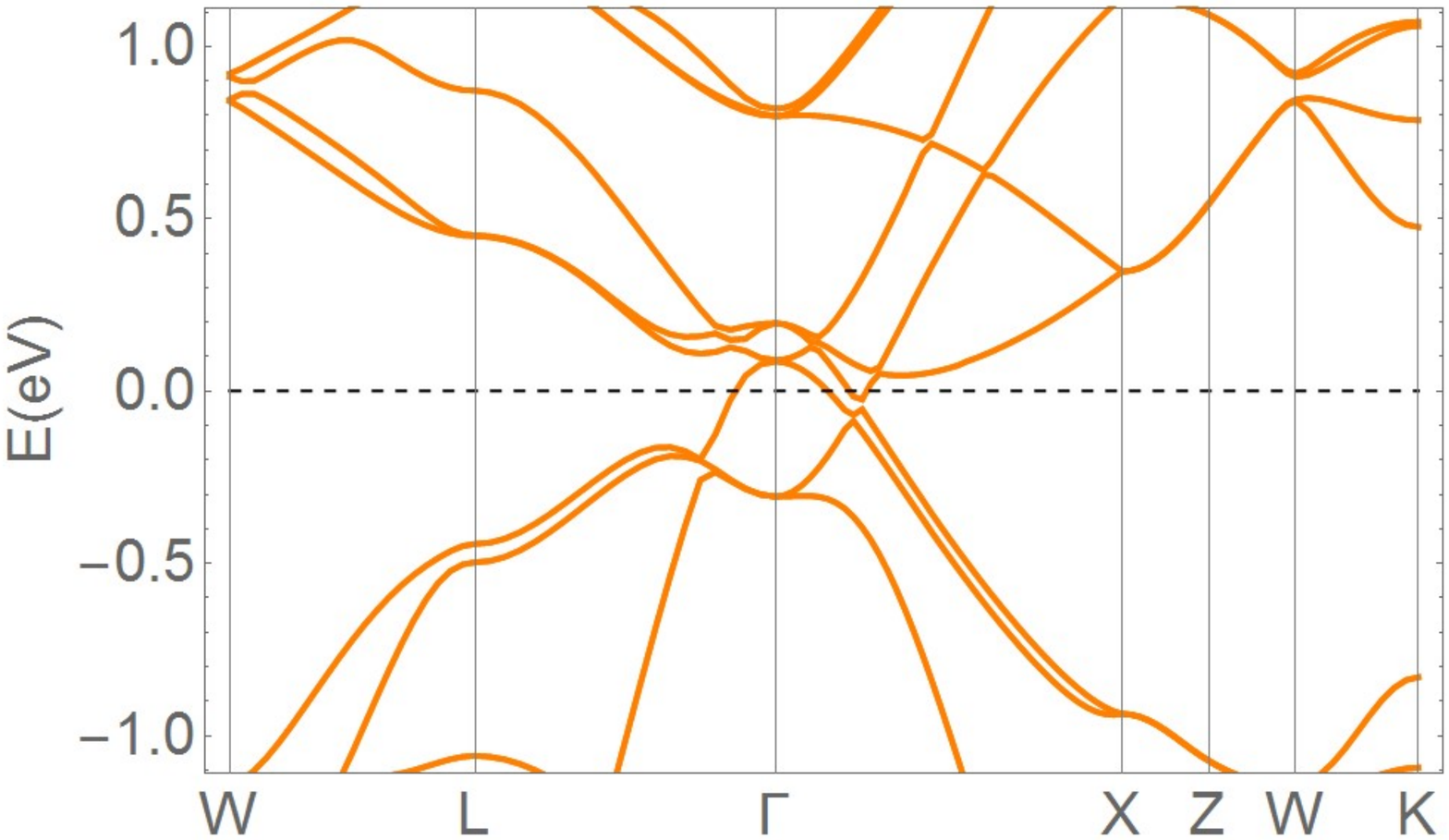}\newline
\caption{The band plot for the hourglass topological insulator c-TiS$_2$. Note that at $\Gamma$ point, there is a finite direct gap.}
\label{app:fig-3}
\end{figure}

As shown in BiBr of the main text, symmetry other than inversion may protect
gapless surface states. In that case, $C_2$ rotation plays such a role while
the screw 2-fold axis in MoTe$_2$ cannot. However the other kind of
nonsymmorphic operation, i.e. glide plane may protect novel hourglass
surface states \cite{hourglass}. We thus search the $\mathcal{SG}\mathbf{227}$ which has more
symmetry operations especially some glide planes and we found that the cubic
TiS$_{2}$ (c-TiS$_{2}$) \cite{TiS2}%
, is 2 in $\mathbb{Z}_4$. For $\mathcal{SG}\mathbf{227}$, there are $8$ AI
basis vectors while only one has common factor (equal to 4) \cite{Po} (shown in Sec. \ref{AIS}), thus
its SI group $X_{\mathrm{BS}}=\mathbb{Z}_4$. For c-TiS$_{2}$, a primitive
unit cell contributes $\nu=64$ valence electrons (For Ti and S, consider 4
and 6 valence electrons respectively while the multiplicity of the chemical
formula units in the primitive unit cell is $4$). Following the SI strategy,
we calculate $n_{\mathbf{k}}^{\alpha }$ for HSPs up to the 64th band. All these numbers are integers. In Fig. \ref{app:fig-3},there is actually a small but finite direct gap at $\Gamma$ point.  The
expansion on the AI basis vectors is $\mathbf{q}=(3,-2,1,0,1,0,0,-\frac{1}{2}%
)$ . Thus ${n_{\mathbf{k}%
}^{\alpha }}^{\prime}s$ can constitute a BS, and furthermore this BS has a
nontrivial SI (i.e. it has 2 in the $\mathbb{Z}_4$ group). Parity
calculations show that $(\nu _{0};\nu _{1},\nu _{2},\nu _{3})$ are all
vanishing. Thus c-TiS$_2$ is not a strong or weak TI. The inversion
topological invariant \cite{Khalaf,Fang-3} i.e. $\kappa_{1}$ is equal to 2. This means
there is opportunity that we observe the gapless hinge states in the domain
wall of two gapped side surfaces. We note that c-TiS$_{2}$ has several glide
planes, thus may possesses novel surface states. Here the hourglass index ($%
\delta _{g_{001}}$ protected by glide plane $(001)$) must be 1 \cite{Fang-3}
thus we expect hourglass surface states to appear in the surface termination
where the above glide symmetry is preserved.

\subsection{Weak topological insulator: Ag$_2$F$_5$}

\begin{figure}[htbp!]
\includegraphics[width=15cm]{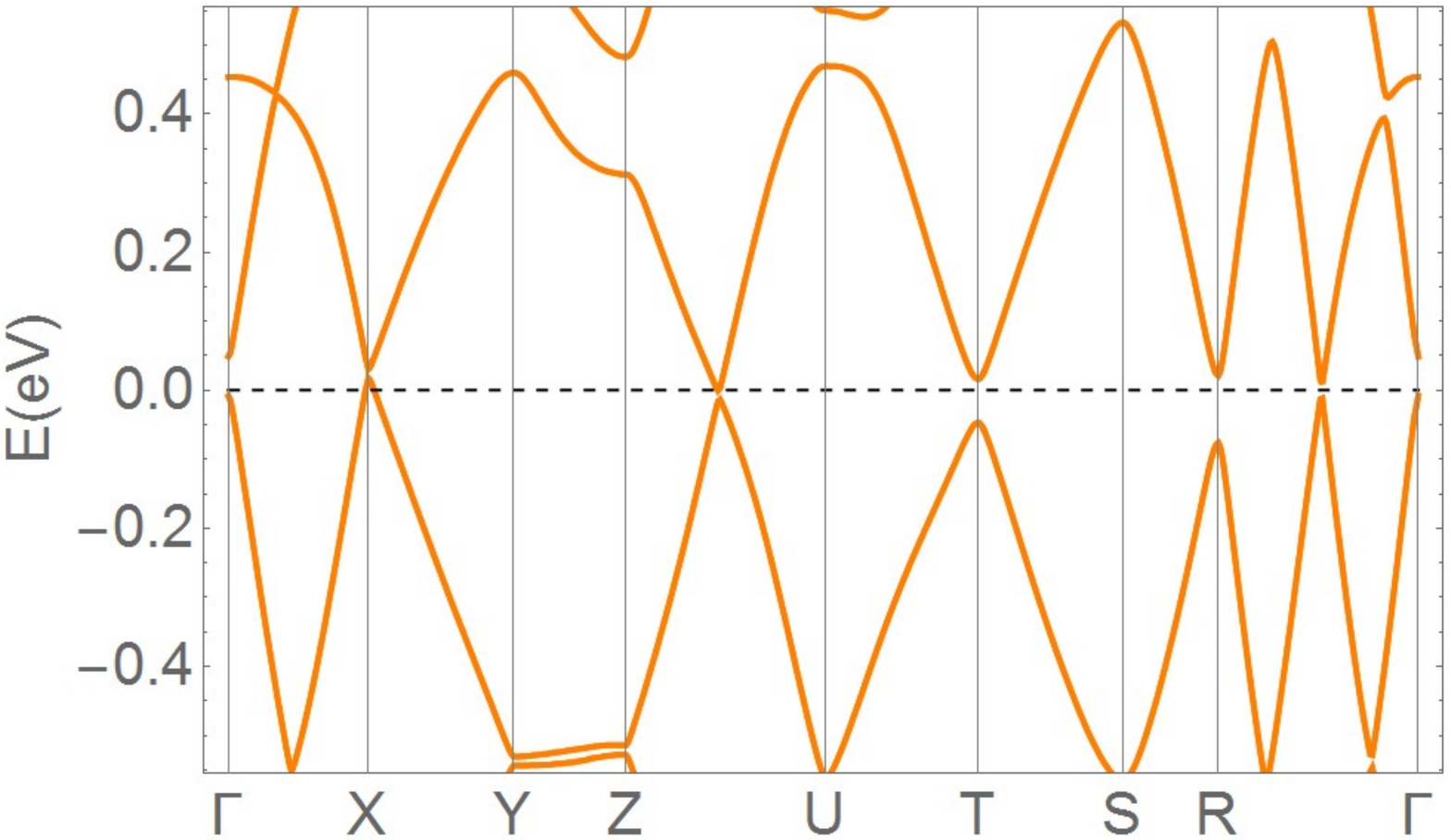}\newline
\caption{The energy band plot for the weak TI for Ag$_2$F$_5$.}
\label{app:fig-2}
\end{figure}

Even with only inversion symmetry, namely $\mathcal{SG}\mathbf{2}$, it can also protect gapless surface states for the compounds
with the SI as 2 in the key $\mathbb{Z}_{4}$. Ag$_{2}$F$_{5}$ is such kind
of material whose ${n_{\mathbf{k}}^{\alpha }}^{\prime}s$ can constitute a BS
indicating the existence of a continuous direct gap throughout the BZ. Its
SI is found to be $(1,1,0,2)$ ($X_{\mathrm{BS}}=\mathbb{Z}_2\times\mathbb{Z}%
_2\times\mathbb{Z}_2\times\mathbb{Z}_4$ \cite{Po}).
As the $\mathbb Z_2$ factors in $X_{\mathrm{BS}}$ here corresponds to weak TIs \cite{Fu-Kane}, this material is a weak TI with translation protected gapless surface states
appearing in (010) and (001) surface. For its 228 valence bands, our first principles calculations show that apart from $\Gamma$ and $X$ points, these 114 KPs at each of the rest TRIM are classified to 57 even and 57 odd KPs. For $\Gamma$ and $X$ points, they both have 66 even and 48 odd KPs. Thus $\kappa_1=2$, and it hosts 1D hinge states. Besides, $\nu_1=1$ and $\nu_{0,2,3}=0$.  By tuning the occupied band parities at $\Gamma$ to 68 even KPs and 46 odd KPs, this compound will become the ``0 in $\mathbb{Z}_4$'' phase as shown in Fig. 2(a). $\kappa_2$ is then equal to 0. So the 1D hinge states will disappear. However $\nu_1$ is still 1, thus the topological surface states still exist \cite{Fu-Kane}.  Seen from the electronic structure
plot shown in Fig. \ref{app:fig-2}, although its full gap is not very large,
 the dispersion looks very clear and so the projection of the
bulk bands to the surface BZ can demonstrate several empty zones for the
accommodation of the surface states. This is very favorable for the
experimental observations.

\subsection{Helical Hinge gapless states in A7 phosphorus}

\begin{figure}[!htbp]
\includegraphics[width=15cm]{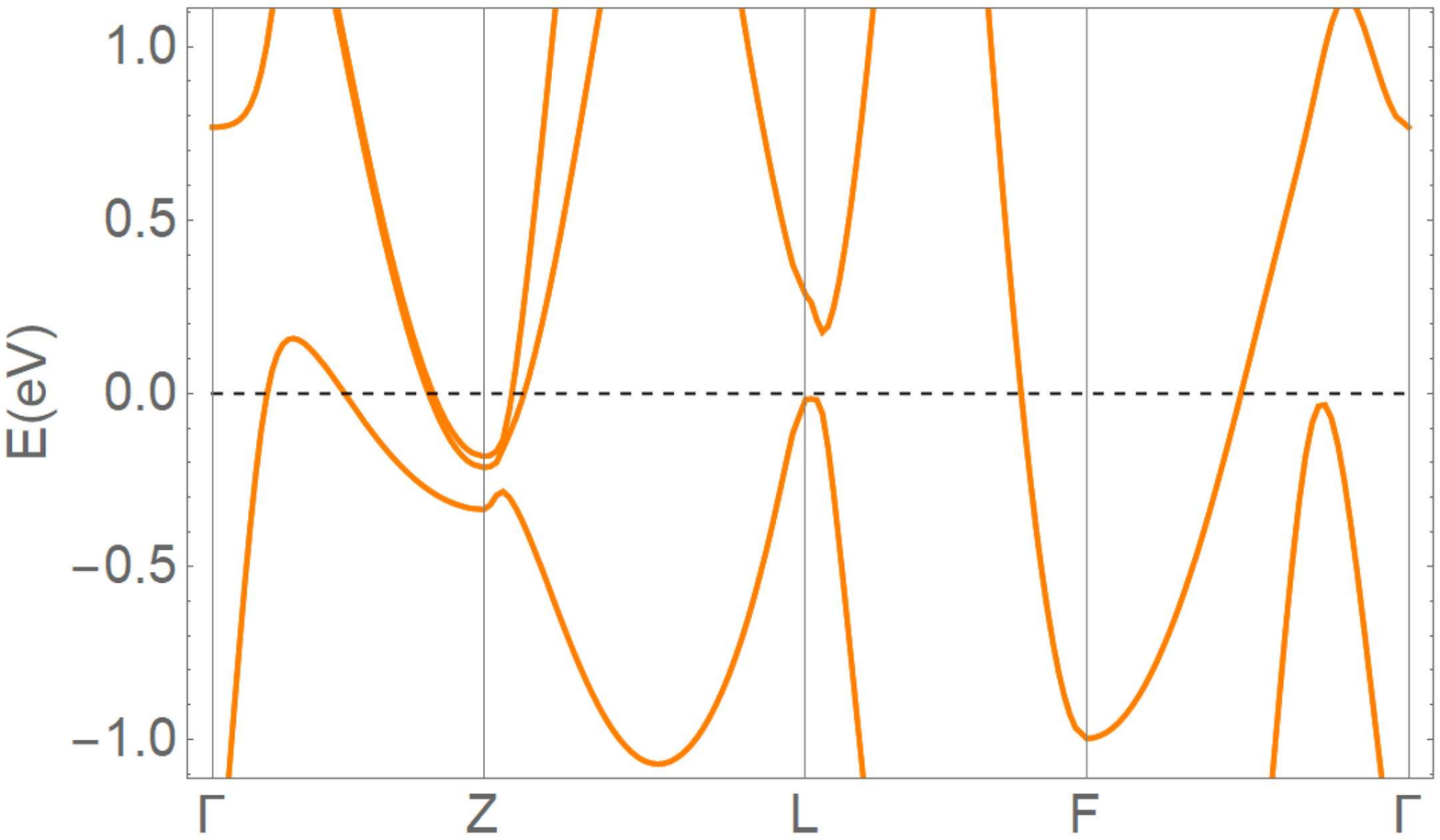}\newline
\caption{The energy band plot of the A7-P.}
\label{app:fig-1}
\end{figure}

It is well-known that the elementary phosphorus (P) owns many kinds of
allotropes such as the Black phosphorus \cite{BP}, a famous layer material. Under
pressure about 9 GPa (whose band plot is shown Fig. \ref{app:fig-1} ), P
crystallizes in $\mathcal{SG}\mathbf{166}$ (A7 phase) \cite{P}. For this $\mathcal{SG}
$, it has in total 8 AI basis vectors while only two have common factor, one
is 4 while the other is 2, thus its SI group is $X_{\mathrm{BS}}=\mathbb{Z}%
_2\times\mathbb{Z}_4$ \cite{Po}. For the A7 P, it has 10 valence electrons
in the primitive unit cell, and we calculate the ${n_{\mathbf{k}}^\alpha}%
^{\prime }s$ for the first 10 bands. The expansion coefficients are
(0,0,1,-1,1,0,-1,$-\frac{1}{2}$) (See Table for the corresponding AI basis
vectors), thus the A7 P, is a BI which has a finite direct band gap
everywhere in the BZ and has a nonvanishing SI=(0,2). Furthermore its
inversion topological invariant is $\kappa_1=2$ \cite{Khalaf,Fang-3}, thus
it must be a topologically nontrivial insulator although all the
conventional topological invariants are found to be vanishing. It displays
inversion-protected gapless hinge states as long as the open conditions
preserve the inversion symmetry. We note that recently there has been an
experimental observation of such hinge states on bismuth \cite{Bi} which shares
the same crystal structure and SI as A7 P here.

\section{Predicted strong topological insulators by $1,3$ in $\mathbb{Z}_4$}
\begin{figure}[!htbp]
\includegraphics[width=20cm]{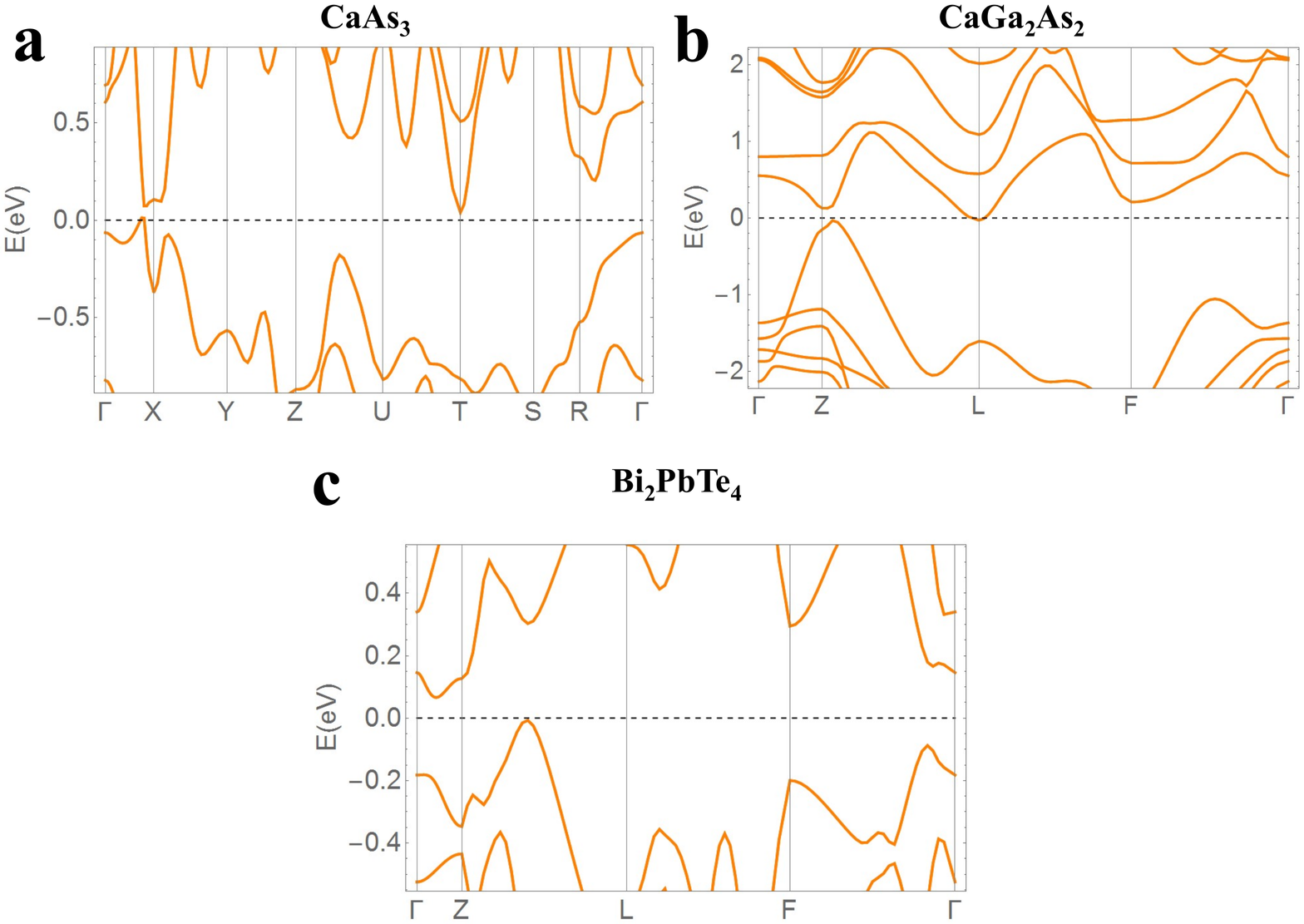}\newline
\caption{The band plots for all the STIs found by ``1 (or 3) in $\mathbb{Z}_4$''.}
\label{app:fig-sti}
\end{figure}

We also find several materials having ``1 (or 3) in $\mathbb{Z}_4$'' which must be strong TIs.  Their electronic band plots are gathered in Fig. \ref{app:fig-sti}. These materials all own  finite direct gap everywhere in the whole BZ, and for Bi$_2$PbTe$_4$, it has a full gap ($\sim 66$ meV).
\begin{table*}[!htb]
\caption{Table of noncentrosymmetric STI candidates discovered by $1,3$ in $\mathbb{Z}_4$ . }
\label{tab:cen-stiwti}%
\begin{tabular}{|c|c|c|c|c|c|}
\hline\hline
$\mathcal{SG}$ & Material &  $X_{\mathrm{BS}}$ & SI & $%
(\nu_0;\nu_1,\nu_2,\nu_3)$ & $\kappa_1$ \\ \hline
\textbf{2} & CaAs$_3$ \cite{caas3} &  $\mathbb{Z}_2\times\mathbb{Z}_2\times\mathbb{Z}%
_2\times\mathbb{Z}_4$ & (0,0,1,1) & (1;1,0,0) & 1 \\ \hline
\textbf{166} & Bi$_2$PbTe$_4$ \cite{bi2pbte4}  & $\mathbb{Z}_2\times\mathbb{Z}_4$ &
(1,1) & (1;1,1,1) & 3 \\ \hline
\textbf{166} & CaGa$_2$As$_2$ \cite{caga2as2} &  $\mathbb{Z}_2\times\mathbb{Z}_4$ &
(1,1) & (1;1,1,1) & 1 \\ \hline
\end{tabular}%
\end{table*}


\section{The other predicted semimetals by SI method}

Through the expansion coefficients $q_i's$, we found another two  topological (semi-)metals in this sections: three-fold degenerate fermions in AuLiMgSn and  nodal-line semimetal AgF$_2$.

\subsection{Three-fold degenerate fermions}

\begin{figure*}[!htbp]
\includegraphics[width=15cm]{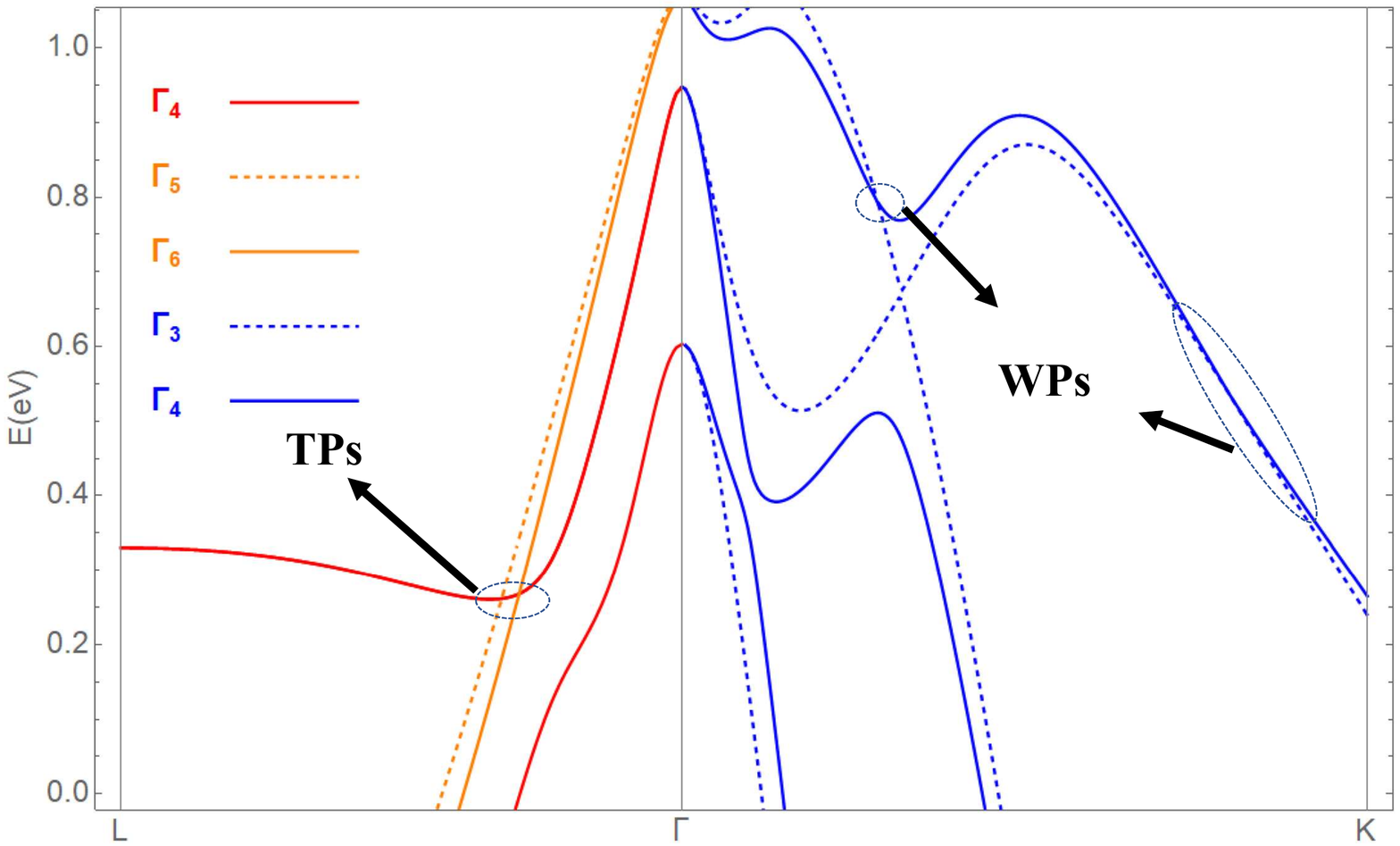}\newline
\caption{ We show the band plot for the AuLiMgSn. Only $%
\Gamma_4$ in the $\Gamma L$ line are 2D irrep and the others are all 1D.
Thus in $\Gamma L$ line, the $\Gamma_4$ band crosses with the $\Gamma_5$ and
$\Gamma_6$, resulting in two kinds of three-fold degenerate points (TPs) or
three-fold degenerate fermions, protected by $C_{3v}$. In $\Gamma K$ line, $%
\Gamma_3$ and $\Gamma_4$ crosses with each other resulting in two type-II
Weyl points (WPs), which are protected by $S_2$ symmetry. }
\label{app:fig-4}
\end{figure*}

The three-fold degenerate fermions are found for AuLiMgSn \cite{aulimgsn} which has $%
\mathcal{SG}216$. The calculated ${n_{\mathbf{k}}^\alpha}^{\prime }s$ are
all integers thus there are finite direct gaps in all the HSPs. However
expansion on the $\mathcal{SG}216^{\prime }s$ AI basis vectors shows that
they cannot constitute a BS at all, namely case 3 in the main text. Thus there must be some band crossing.
Based on the first principles calculations , we  find that in the symmetry line $\Gamma L$, the $\Gamma_4$ and $\Gamma_6$ bands shown in Fig. \ref{app:fig-4} cross with each other,
resulting in a three-fold degenerate fermion (TF). Meanwhile near such a
band crossing, there is also another TF originated from the crossing between
$\Gamma_4$ and $\Gamma_5$. These TFs are protected by $C_{3v}$ symmetry
along the $\Gamma L$. Furthermore, in the symmetry line $\Gamma K$ whose
symmetry group is $S_2$, we also observe two band crossings between two
nondegenerate bands, i.e. resulting in two Weyl points (WPs). These WPs carry a zero topological charge due to $S_2$. They are also shown in Fig. %
\ref{app:fig-4}.

\subsection{The nodal-line semimetal  AgF$_2$}

\begin{figure}[htbp]
\includegraphics[width=20cm]{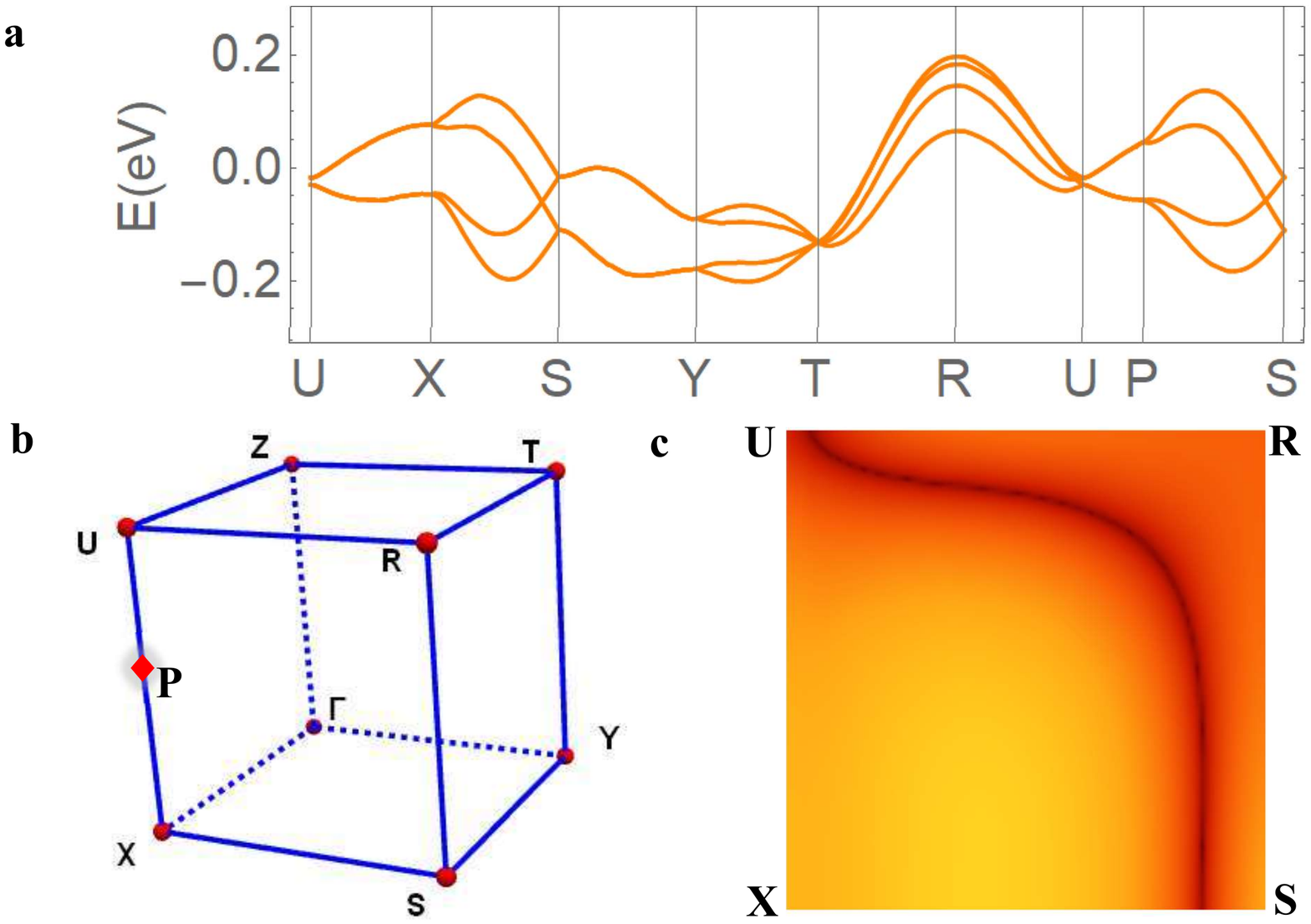}\newline
\caption{ (a) shows the energy band plot of AgF$_2$. (b)
depicts the $\frac{1}{8}$ Brillouin zone in the left panel. In the up right
panel of (b), from $S$ to any point $P$ in $UX$, there is always a hourglass
type band crossing. This then results a nodal line as shown in the down
right panel of (b) based on the first principles calculations.}
\label{app:fig-5}
\end{figure}

AgF$_{2}$  \cite{agf2}($\mathcal{SG}\mathbf{61}$) is predicted as a topological bulk
hourglass nodal-line semimetal. In AgF$_{2}$, there are in total 60 valence
electrons. Thus $\nu_{e}=60$. We calculate the numbers ${n_{\mathbf{k}%
}^{\alpha }}^{\prime }s$ up the 60th bands and find that they are all
integers, which means that at least at HSPs the valence bands are gapped
from the conduction bands. However they cannot constitute a BS at all
because the expansion is $(\frac{11}{2},\frac{19}{2},\frac{13}{4})$ (See Sec. \ref{AIS} for the AI basis vectors of $\mathcal{SG}\mathbf{61}$). Hence there must be
some band crossing(s) in the BZ. Inspecting all the high symmetry lines and
planes, we find that in $k_{x}=\frac{\pi }{a}$ plane, there is a large
four-fold-degenerate nodal-line shown in Fig. \ref{app:fig-5}(c) by first principles calculations. The glide
symmetry, i.e. $\tilde{M}_{x}=(-x+\frac{1}{2},y+\frac{1}{2},z)$, will guarantee that
any curve connecting $S$ and $P$ (arbitrary point in $UX$) will possess an
unavoidable hourglass type band crossing (See Fig. \ref{app:fig-5}(a)). The crossing point is robust and
protected by $\tilde{M}_{x}$ because it is originated from two 2-fold degenerate
bands with inverse eigenvalue of $\tilde{M}_{x}$. These crossing points form a hourglass Dirac nodal line in Fig. \ref{app:fig-5}(c).

The $\mathcal{SG}\mathbf{61}$ is a nonsymmorphic group with two glides: $\tilde{M}_x=(-x+\frac{1}{2},y+\frac{1}{2},z)$, $\tilde{M}_y=(x,-y+\frac{1}{2},z+\frac{1}{2})$  and
inversion $\mathcal{I}$. A third glide plane can be obtained by the product of the
above two: $\tilde{M}_z=(x+\frac{1}{2},y,-z+\frac{1}{2})$. In the $U$-$X$ line i.e. $%
k_x=\pi,k_y=0$, the symmetry includes $\tilde{M}_x$ and $\tilde{M}_y$, $\mathcal{TI}$ and their products.
We first consider the first two operations. The eigenvalues of them can be
quickly obtained through: $\tilde{M}_x^2=\{\bar{E}%
|0,1,0\}=-1$ and $\tilde{M}_y^2=\{\bar{E}%
|0,0,1\}=-e^{-ik_z}$ where $\bar{E}$ represent spin-$2\pi$ rotation. Thus
the eigenvalues for the two can be $g_x=\pm i$ and $g_y=\pm ie^{-i\frac{k_z}{%
2}}$. Besides they anticommutates with each other, therefore we can only use
$g_x$ or $g_y$ to label the Bloch eigen-states. We choose $g_x$ here and
label the Bloch states as $|\mathrm{UX},g_x>$, and $\tilde{M}_y|\mathrm{UX},g_x>=e^{-ik_z}|\mathrm{UX},-g_x>$. Hence $|\mathrm{%
UX},g_x>$ and $|\mathrm{UX},-g_x>$ will share the same eigen energy (they
are orthogonal to each other because of the inverse $g_x$).\newline

It is well-known that $\mathcal{TI}$  enforces each band to be at least
two-fold degenerate. And because ${%
\mathcal{I},\tilde{M}_x}=0$, $\mathcal{TI}$ will preserve the
eigenvalue $g_x$, i.e., $\tilde{M}_x(\mathcal{TI}|%
\mathrm{UX},g_x>)=g_x(\mathcal{TI}|\mathrm{UX},g_x>)$. This will require the
aforementioned doublet to be doubled, which results a 4D irrep \cite{Bradley}%
. \newline

We then consider the symmetry line $S$-$X$, i.e. $k_x=\pi, k_z=0$. Then we should
consider $\tilde{M}_x$, $\tilde{M}_z$, $\mathcal{TI}$ and their products. As before, we first consider the
eigen-values of the first two operations in this line. Because $\tilde{M}_x^2=-e^{-ik_y}$ and $\tilde{M}_z^2=1 $, then the eigenvalues have to be $g_x=\pm i e^{-i\frac{k_y}{2}}$
and $g_z=\pm 1$ respectively. Furthermore they commutates with other, thus
we can label the Bloch eigen states by the combination of their eigenvalues
i.e. $|\mathrm{SX},g_x,g_z>$. Actually there are 4 1D irrep's in this line
\cite{Bradley}, \newline

Again we consider $\mathcal{TI}$. First$\tilde{M}_x\mathcal{I}=\{E|1,1,0\}\mathcal{I}\tilde{M}_x=-e^{-ik_y}\tilde{M}_x$. This means that $\mathcal{TI}$ will preserve $g_x$.
Similarly, $\mathcal{TI}$ will reverse $g_z$. Hence, in $S$-$X$, two-fold (due to
$\mathcal{TI}$) degenerate bands bear the same $g_x$ and inverse $g_z$. In
another words in $S$-$X$, there are two different 2D irreps when considering time
reversal symmetry, which will converge into a 4D irrep at $S$ or two 4D
irrep's at $X$ \cite{Bradley}. Thus it is possible for the band crossing to
happen in $S$-$X$.
$S$ and $X$ are both time-reversal invariant. And $P$ commutates or
anti-commutates with $\tilde{M}_x$ for $S$ or $X$
respectively, while $P$ anti-commutates with $\tilde{M}_z$ both for $S$ and $X$. At $S$, $\mathcal{P}|\mathrm{S},g_x,g_z>$ bears the same
eigenvalue $g_x$ while inverse eigenvalue $g_z$ as $|\mathrm{S},g_x,g_z>$. $%
\mathcal{T}$ will not change both $g_x$ and $g_z$, thus at $S$ the quartet $|%
\mathrm{S},g_x,g_z>,\mathcal{I}|\mathrm{S},g_x,g_z>,\mathcal{T}|\mathrm{S},g_x,g_z>,%
\mathcal{TI}|\mathrm{S},g_x,g_z>$ bears the same $g_x$ and are mutually
orthogonal. Actually S bears two 2D irrep's both are doubled by $\mathcal{T}$
\cite{Bradley}. However, at $X$, it is similar to find that $|\mathrm{X}%
,g_x,g_z>,\mathcal{I}|\mathrm{X},g_x,g_z>,\mathcal{T}|\mathrm{X},g_x,g_z>,\mathcal{TI}|%
\mathrm{X},g_x,g_z>$ have $g_x=(i,-i,-i,i)$ or $(-i,i,i,-i)$. Because the
eigenvalue $g_x$ continuously changes by $g_x=\pm ie^{-i\frac{k_y}{2}}$,
there must be states switch between the quartets of $S$ and $X$ through $S$-$X$,
which results in a hourglass pattern and a Dirac point. Further as $g_x$ is
well-defined in the whole line $U$-$X$, any curve connecting $S$ and one point in
$U$-$X$ would give a hourglass type Dirac band crossing. This is verified by our
first principles calculations shown in Fig. \ref{app:fig-5}(b). This is why
we cannot obtain a BS when consider bands up to the filling.

\section{First principles calculations of mirror Chern numbers}
\subsection{Techniques involved in first principles calculations of mirror Chern numbers}
The central task to calculate the (mirror) Chern number is  to calculate the overlap matrix $S_{nn^{\prime }}(\mathbf{k},%
\mathbf{k}^{\prime })=<u_{n\mathbf{k}}|u_{n^{\prime }\mathbf{k^{\prime }}}>$
where the inner product is the integration in the primitive unit cell and $u_{n\mathbf{k}}$ is the Bloch eigen-state. The
mirror Chern number, namely $C_M$ is defined in the mirror symmetric plane in the BZ: $C_M=%
\frac{1}{2}(C_M^{+i}-C_M^{-i})$, where the superscript $\pm i$ labels the
eigenvalues of the mirror operation. Due to $\mathcal{T}$, $C_M^{+i}=-C_M^{-i}$, thus $C_M=C_M^{+i}$.  By definition:
\begin{equation}  \label{app:chern}
C_{M}^{\pm i}=\sum_{n\in occ.}\int_{BZ}d^2\mathbf{k}\Omega^{\pm i}(n\mathbf{k%
}),
\end{equation}
where $occ.$ denotes the occupied bands and $\Omega^{\pm i}(n\mathbf{k})$ is
the Berry curvature at $\mathbf{k}$ for $|u_{n\mathbf{k}}^{\pm i}>$ where we
have used the mirror eigenvalues to label the lattice-periodic function $%
|u_{n\mathbf{k}}^{\pm i}>$:
\begin{equation}  \label{app:berrycur}
\Omega^{\pm i}(n\mathbf{k})=(\nabla \times \mathbf{A}^{\pm i}(n\mathbf{k}%
))_\perp,
\end{equation}
where $\perp$ means the direction perpendicular to the symmetric plane, and $%
\mathbf{A}$ is the Berry connection: $\mathbf{A}^{\pm i}(n\mathbf{k}%
)=i<u^{\pm i}_{n\mathbf{k}}|\partial_\mathbf{k}|u^{\pm i}_{n\mathbf{k}}>$.
For a small portion $\Delta S $ is the symmetry plane, the following
relation must hold:
\begin{equation}  \label{app:small}
\gamma^{\pm i}_{\Delta S}(n)=\int_{\Delta S}d^2\mathbf{k}\Omega^{\pm i}(n%
\mathbf{k})=\oint_{\partial S} d\mathbf{r}\cdot\mathbf{A}^{\pm i}(n\mathbf{k}%
)\in(-\pi,\pi],
\end{equation}
where $\partial S$ represent the boundary of $\Delta S$ and we have chosen a gauge that the Berry phase is restricted into $%
(\pi,-\pi]$ which is required by that the Berry curvature is finite while
  $\Delta S$ is very small. Because,
\begin{equation}  \label{app:diff}
\mathbf{A}^{\pm i}(n\mathbf{k})=i\frac{<u^{\pm i}(n\mathbf{k})|u^{\pm i}(n,%
\mathbf{k}+\delta\mathbf{k})>-<u^{\pm i}(n\mathbf{k})|u^{\pm i}(n,\mathbf{k}%
)>}{\delta \mathbf{k}},
\end{equation}
we will have,
\begin{equation}  \label{app:overlap}
<u^{\pm i}(n\mathbf{k})|u^{\pm i}(n,\mathbf{k}+\delta\mathbf{k})>=e^{-i%
\mathbf{A}^{\pm i}(n\mathbf{k})\cdot\delta\mathbf{k}}.
\end{equation}
Thus we can divide the loop $\partial S$ into several parts: $(\mathbf{k}_1,%
\mathbf{k}_2]\cup(\mathbf{k}_2,\mathbf{k}_3]\cup\ldots\cup(\mathbf{k}_{N-2},%
\mathbf{k}_{N-1}]\cup(\mathbf{k}_{N-1},\mathbf{k}_1]$. According to Eq. (\ref%
{app:overlap}),
\begin{equation}  \label{app:berryphase}
\gamma^{\pm i}_{\Delta S}(n)=-\mathrm{Im}log(<u^{\pm i}_{n\mathbf{k}%
_1}|u^{\pm i}_{n\mathbf{k}_2}><u^{\pm i}_{n\mathbf{k}_1}|u^{\pm i}_{n\mathbf{%
k}_2}>\ldots<u^{\pm i}_{n\mathbf{k}_{N-1}}|u^{\pm i}_{n\mathbf{k}_1}>).
\end{equation}
It is then easy to generalize to the multi-band case:
\begin{equation}  \label{app:multi-berryphase}
\sum_{n\in occ.}\gamma^{\pm i}_{\Delta S}(n)=-\mathrm{Im}log \mathrm{det}[S^{\pm i}(%
\mathbf{k}_1,\mathbf{k}_2)S^{\pm i}(\mathbf{k}_1,\mathbf{k}_2)\ldots S^{\pm
i}(\mathbf{k}_{N-1},\mathbf{k}_1)],
\end{equation}
where det is short for determinant , $S$ is the overlap matrix
whose row and column indices are the occupied band indices $n,n'$, $S$ is $%
\nu\times\nu$ assuming there are in total $\nu$ occupied bands and the
superscripts $\pm i$ labels the eigenvalue of mirror as before. Eq. (\ref{app:multi-berryphase}) is clearly gauge independent for the simultaneous
presences in the bra and ket for a eigen-function. Note that the
first principles eigen-functions  generally are not simultaneously
the eigenstates of the mirror operator, thus a transformation should be made to
obtain the eigenstates of the mirror, denoted as $U_m(\mathbf{k}_1)$ as $%
\mathbf{k}_1$ with the first half with eigenvalue $+i$ and the rest with
eigenvalue $-i$. Write Eq. (\ref{app:multi-berryphase}) as follows:
\begin{eqnarray}  \label{app:whole-berryphase}
\sum_{n\in occ.}\gamma_{\Delta S}(n)=-\mathrm{Im}log|S(\mathbf{k}_1,\mathbf{k%
}_2)S(\mathbf{k}_1,\mathbf{k}_2)\ldots S(\mathbf{k}_{N-1},\mathbf{k}_1)| \\
=-\mathrm{Im}log|U_m(\mathbf{k_1})^\dag S(\mathbf{k}_1,\mathbf{k}_2)S(%
\mathbf{k}_1,\mathbf{k}_2)\ldots S(\mathbf{k}_{N-1},\mathbf{k}_1)U_m(\mathbf{%
k_1})| \\
=-\mathrm{Im}log|[U_m(\mathbf{k_1})^\dag S(\mathbf{k}_1,\mathbf{k}_2)S(%
\mathbf{k}_1,\mathbf{k}_2)\ldots S(\mathbf{k}_{N-1},\mathbf{k}_1)U_m(\mathbf{%
k_1})]_{1\sim\nu/2,1\sim\nu/2}| \\
-\mathrm{Im}log|[U_m(\mathbf{k_1})^\dag S(\mathbf{k}_1,\mathbf{k}_2)S(%
\mathbf{k}_1,\mathbf{k}_2)\ldots S(\mathbf{k}_{N-1},\mathbf{k}_1)U_m(\mathbf{%
k_1})]_{\nu/2\sim\nu,\nu/2\sim\nu}|,
\end{eqnarray}
where in the last equality, the two parts correspond to $\sum_n\gamma_{%
\Delta S}(n)^{+i}$ and $\sum_n\gamma_{\Delta S}(n)^{-i}$ respectively.
Finally, to calculate the mirror Chern number, we just need to calculate the
overlap matrix $S(\mathbf{k},\mathbf{k}^{\prime })$ no matter whether the
eigenfunctions involved are the eigen-states of mirror or not and we just
need to make a unitary transformation shown above to extract the parts for
each mirror-eigenvalue.

\subsection{Details for calculation of $C_M$ in $\beta$-MoTe$_2$ and BiBr}

\begin{figure*}[tbp]
\includegraphics[width=20cm]{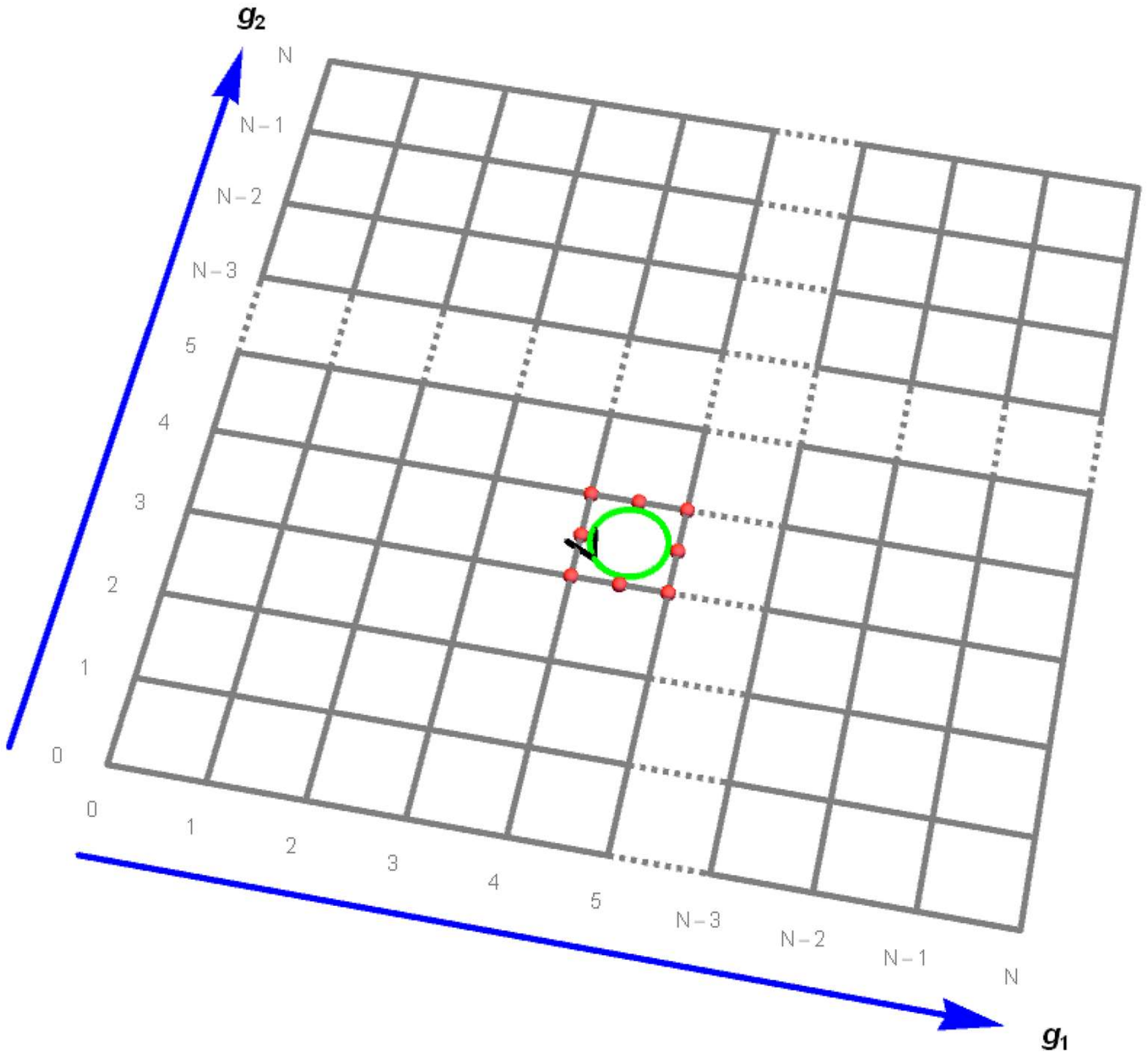}\newline
\caption{The sketch for numerical calculation of the Chern number. First we dived the 2D BZ into $N\times N$ small portions. For each portion $\Delta S$, we then calculate the Berry phase around its boundary $\partial S$ through further dividing it into several segments (red dots).}
\label{app:Chern}
\end{figure*}

For $\beta$-MoTe$_2$,  the mirror plane is perpendicular to  the $z$(or $\mathbf{c}$) direction: $M_z=(x,y,-z+\frac{1}{2})$, we should focus on two mirror symmetric
planes $k_z=0$ and $k_z=\frac{\pi}{c}$ respectively. Note that in $k_z=\frac{%
\pi}{c}$ plane, $\mathcal{TI}$ will preserve the mirror eigen-values due to the $\frac{1}{2}\mathbf{c}$ translation in $M_z$: i.e.
when $\psi$ is the eigenstate of $M_z$, $\mathcal{TI}\psi$ will have the
same eigenvalue of $M_z$. Not that $\mathcal{TI}$ will change the sign of
the Berry curvature, thus the mirror Chern will be vanishing for the $k_z=%
\frac{\pi}{c}$ plane. For the $k_z=0$ plane, we exploit the  above technique through linearized augmented plane-wave method
as implemented in WIEN2k package \cite{wien2k}. We divide the $k_z=0$ plane to to $%
50\times50$ parts and for each part we calculate the Berry phase around
them (Eq. (\ref{app:whole-berryphase})) by dividing each part into 8
portions (schematically shown in Fig .\ref{app:Chern}). The calculated MCNs are found to be vanishing. According to the formula of $C_M^+=i\sum_{n}\int d^{2}k<\partial
_{k_{x}}u_{n}^{+}|\partial _{k_{y}}u_{n}^{+}>-c.c.$ where $+$ represents that
the eigenvalue of the mirror operator is $+i$ , $n$ is the occupied band
index and the integral zone is restricted to a mirror symmetric plane. Our
calculations find that the eigen-state $u_{n}^{+}$ has a weak dependence on $%
k_{x}$ which is consistent with that the MCNs are all vanishing.

 For BiBr, the mirror plane is also perpendicular to $\mathbf{c}$ in our adopted setting. In this case, $\mathcal{TI}$ will reverse the eigenvalue of the mirror operator. So we should calculate the mirror Chern numbers for both $k_z=0$ and $k_z=\frac{\pi}{c}$ planes. We take the similar way of partitioning the 2D BZ  as $\beta$-MoTe$_2$.

\section{Tight binding model for M\lowercase{o}T\lowercase{e}$_2$ and B\lowercase{i}B\lowercase{r}}\label{TB}

For the calculations of the hinge states for MoTe$_2$ and BiBr, we need to
take open conditions in two directions. It would be rather computationally
demanding especially for the first principles calculation. Thus we construct
a tight binding (TB) model for both materials:
\begin{equation}  \label{app:Htb}
\hat{H}_{TB}=\sum_{\mathbf{R},\mathbf{R^{\prime }};s,s^{\prime
};\mu,\mu^{\prime };\sigma,\sigma^{\prime }}h(\mathbf{R},\mathbf{R^{\prime }}%
;s,s^{\prime };\mu,\mu^{\prime };\sigma,\sigma^{\prime })\hat{C}^\dag_{%
\mathbf{R}+\boldsymbol{\tau}_s,\mu,\sigma}\hat{C}_{\mathbf{R}^{\prime }+%
\boldsymbol{\tau}_{s^{\prime }},\mu^{\prime },\sigma^{\prime }},
\end{equation}
where $\mathbf{R},\mathbf{R^{\prime }}$ label the primitive unit cell within
which the atoms are located at $\boldsymbol{\tau}_s,\boldsymbol{\tau}%
_{s^{\prime }}$ relative the primitive unit cell. $\mu,\mu^{\prime }$ label
the orbital degree of freedom and $\sigma,\sigma^{\prime }$ label the spin
eigenvalue of $S_{z}$ (the $z$ component of the spin operator $\mathbf{S}$).
$\hat{C}^\dag$ ($\hat{C}$) is the creation (annihilation) operator of the
state as denoted in its subscript. We adopt orthogonal atomic orbitals. In
each atom, we choose appropriate atomic orbitals which dominate the
contributions near the Fermi level.  Symmetry
(time-reversal and space group) imposes restrictions for the Hamiltonian
matrix elements. We thus take the Slater-Koster (SK) formalism wherein the
hopping integrals are given by several adjustable parameters and the hopping
direction cosines. While for the onsite terms, the crystal field splitting
can be described by the site-symmetry allowed onsite Hamiltonian matrix.
The spin-orbit coupling (SOC) is given by,
\begin{equation}  \label{app:soc}
\hat{H}_{SO}=\sum_{\mathbf{R}s,\mu\nu,\sigma \sigma^{\prime }}(\lambda_{s}%
\mathbf{L}\cdot\mathbf{S})_{\mu \sigma,\nu \sigma^{\prime }}\hat{C}_{\mathbf{%
R}+\boldsymbol{\tau}_s,\mu,\sigma}^\dag \hat{C}_{\mathbf{R}+\boldsymbol{\tau}%
_s,\nu,\sigma^{\prime }},
\end{equation}
where $\mathbf{L}$ is the orbital momentum and $\lambda_{s}$ is the SOC
parameter: for those atoms related by $\mathcal{SG}$ symmetry, they share
the same SOC parameter while note that $\lambda_s^{\prime }s$ also take a
different values for different $l^{\prime }s$.

With the above TB model at hand, we then fit the energy bands from the first
principles calculation near the Fermi level within the
irreducible BZ for both materials, by least squares method to obtain a
optimized parameters. The comparisons of the TB electronic bands with the first principles results in Fig. \ref{app:fit} show that for both materials, the TB models reasonably reproduce the first principles bands.  Besides the energy bands, it is also required to reproduce
exactly the same SI and mirror Chern numbers as the first principles results.

\section{Method}
The electronic band structure calculations have been carried out using the
full potential linearized augmented plane-wave method as implemented in
the WIEN2K package \cite{wien2k}. The generalized gradient approximation (GGA) with Perdew-Burke-Ernzerhof (PBE) \cite{GGA} realization was adopted for the exchange-correlation functional. It is worth pointing out that  the modified Becke-Johnson exchange potential for the correlation potential (MBJ) \cite{mbj} has also been used and we found that it has no affect on our main results.

\section{Calculation of atomic insulator basis}\label{AI}

The AI basis vectors are the central objects  of our SI method for screen the materials database to  search for topological materials.
Step by step we show how to determine the AI basis for a $\mathcal{SG}$, taking $\mathcal{SG}\mathbf{2}$ as the example:\\

\emph{\textbf{Step 1}}:\\
\textit{Obtaining HSPs.} For $\mathcal{SG}\mathbf{2}$, the HSPs are just 8 time-reversal
invariant momenta (TRIM): $\mathbf{k}=(k_1,k_2,k_3)$ written in the basis of
the reciprocal lattice basis vectors and $k_i$ takes 0 or $\frac{1}{2}$
shown in Table \ref{app:SG2hsp}.
\begin{table}[htbp!]
\centering
\begin{tabular}{|c|c|c|c|c|c|c|c|c|}
\hline
$\mathcal{SG}2^{\prime }s$ HSP & $\Gamma$ & $X$ & $Y$ & $Z$ & $U$ & $T$ & $S$
& $R$ \\ \hline
coordinate & (0,0,0) & $(\frac{1}{2},0,0)$ & $(0,\frac{1}{2},0)$ & $(0,0,%
\frac{1}{2})$ & $(\frac{1}{2},\frac{1}{2},0)$ & $(0,\frac{1}{2},\frac{1}{2})$
& $(\frac{1}{2},0,\frac{1}{2})$ & $(\frac{1}{2},\frac{1}{2},\frac{1}{2})$ \\
\hline
\end{tabular}%
\caption{The coordinates of eight high symmetry points (HSPs) for $\mathcal{%
SG}$2 in the reciprocal lattice basis vectors.}
\label{app:SG2hsp}
\end{table}
\textit{The irreps for each HSP.}\\
For each high symmetry point $\mathbf{k}$,
the little group $\mathcal{G}(\mathbf{k})$ is also $\mathcal{SG}\mathbf{2}$
containing only two different irreps: $D^1(\mathbf{k})$ and $D^2(\mathbf{k}%
) $ that:
\begin{equation}  \label{app:SG2irrep}
\{p|\mathbf{R}\}{\psi_{\mathbf{k},i}^\alpha}=\sum_{i'}\mathrm{D}^\alpha_{\mathbf{k}%
,i^{\prime }i}(\{p|\mathbf{R}\})\psi_{\mathbf{k},i^{\prime }}^\alpha,
\end{equation}
where $\{p|\mathbf{R}\}\in\mathcal{G}(\mathbf{k})$, $\mathrm{D}^\alpha_{%
\mathbf{k},i^{\prime }i}(\{p|\mathbf{R}+\mathbf{R}^{\prime -i\mathbf{b}\cdot%
\mathbf{R}^{\prime }}\mathrm{D}^\alpha_{\mathbf{k},i^{\prime }i}(\{p|\mathbf{%
R}\})$,$\alpha$ labels the irrep taking 1 or 2, and $i,i^{\prime }$ denote
the basis vector for the irrep.\newline

To clearly demonstrate an irrep, we will use the subscript $i$ ($i=1,2,\ldots,r$, $r$ is the total number of the irreps) added to the name of the HSP to
label the $i$th irrep of this HSP. Note that some irrep can be doubled by the
time-reversal operation ($\mathcal{T}$), which means that this irrep must
occur even times, then we should divide the correspond number $n_{\mathbf{k}%
}^\alpha$ by 2\cite{Po}.  In this case, the irrep is in {\color{red}red}.
The dimension of the irrep will be denoted by the superscript added to the HSP's name, e.g. $\Gamma_1^1,\Gamma_2^1,{\color{red}\Gamma_3^1},\Gamma_4^2,\ldots$ represents
the irreps at $\Gamma$ (we will demonstrate the irreps hereafter following
the same order shown in Ref. \cite{Bradley}): $\Gamma_1$ and $\Gamma_2$ are
both 1D irreps while $\Gamma_3$ is 1D irrep doubled by $\mathcal{T}$ and $%
\Gamma_4$ is 2D irrep.\newline

 For $\mathcal{SG}\mathbf{2}$ here, $\mathrm{D}^\alpha(\mathcal{I})=(-1)^{\alpha-1}$,
i.e. $\alpha=1$ (the first irrep) corresponds to the states with even parity while $\alpha=2$ (the second irrep)
corresponds to the states with odd parity. That is to say,   every  HSP of $\mathcal{SG}\mathbf{2}$ has two 1D irrep (each one is doubled by $\mathcal{T}$). So the total number of ${n_{\mathbf{k}}^\alpha}'s$ is 16.
Consider the filling number $\nu$, there are 17 entries in any BS with $\mathcal{SG}\mathbf{2}$. Consider  eight compatibility relations  $\nu=n_{\mathbf{k}}^1+n_{\mathbf{k}}^2$.  Then  only 9 entries  are independent, i.e. $d_{\mathrm{BS}}=9$.  For
an arbitrary $\mathcal{SG}$, the compatibility relations can get much more
complex. However instead  of directly analyzing the compatibility relations,  we can detour to obtaining the AI basis vectors  \cite{Po}. Any AI,
itself a BS, will automatically satisfy the requirement of the compatibility
relations. To obtain the AI basis vectors, one need to exhaustively consider all the Wyckoff positions and all the site-symmetry irreps. \newline

\emph{\textbf{Step 2}}:
\textit{Give all the Wyckoff positions.}  The Wykcoff positions for $\mathcal{%
SG}\mathbf{2}$ is shown in Table \ref{app:sg2wyck}.
\begin{table}[htbp!]
\centering
\begin{tabular}{|c|c|c|}
\hline
$\mathcal{SG}2^{\prime }s$ Wyckoff position & site-group & Wyckoff orbits \\
\hline
2i & $C_1$ & $(x,y,z),(-x,-y,-z)$ \\ \hline
1h & $C_i$ & $(1/2,1/2,1/2)$ \\ \hline
1g & $C_i$ & $(0,1/2,1/2)$ \\ \hline
1f & $C_i$ & $(1/2,0,1/2)$ \\ \hline
1e & $C_i$ & $(1/2,1/2,0)$ \\ \hline
1d & $C_i$ & $(1/2,0,0)$ \\ \hline
1c & $C_i$ & $(0,1/2,0)$ \\ \hline
1b & $C_i$ & $(0,0,1/2)$ \\ \hline
1a & $C_i$ & $(0,0,0)$ \\ \hline
\end{tabular}%
\caption{The nine Wyckoff positions for $\mathcal{SG}$$\mathbf{2}$.}
\label{app:sg2wyck}
\end{table}

\textit{List all the site-symmetry irreps for every Wyckoff position.}Given a Wyckoff position $m\mathcal{W}$ (like $%
2a,4b,\ldots$, $m$ counts the number of sites in this Wyckoff orbit), its
site symmetry group is  then determined. Writing the sites as $\{\mathbf{r}_1^\mathcal{W},\mathbf{r}^\mathcal{W}%
_2,\ldots,\mathbf{r}^\mathcal{W}_m\}$ in one primitive unit cell with the
operations $\{p|\mathbf{R}\}$ in the $\mathcal{SG}$ satisfying $\{p|\mathbf{R%
}\}\mathbf{r}_1^\mathcal{W}=\mathbf{r}_1^\mathcal{W}$ constituting the site
symmetry group. The number of the sites in the Wyckoff orbit $m=\frac{|%
\mathcal{G}_0|}{|\mathcal{G}(\mathbf{r}_1^\mathcal{W})|}$ ($\mathcal{G}_0$
is the point group), and there must exist a $\mathcal{SG}$ element which
will give $\mathbf{r}^\mathcal{W}_J$ ($J=1,2,\ldots,m$) from $\mathbf{r}_1^%
\mathcal{W}$: $\{p_J|\mathbf{R}_J\}\mathbf{r}_1^\mathcal{W}=\mathbf{r}_J^%
\mathcal{W}$. Denoting the basis functions for one irrep ($D_s^\beta$%
of the site symmetry group as $\phi_{\mathbf{r}_1^\mathcal{W},j}^\beta$,
i.e., for $\{p|\mathbf{R}\}\in G(\mathbf{r}^\mathcal{W}_1), \{p|\mathbf{R}%
\}\phi_{\mathbf{r}_1^\mathcal{W},j}^\beta=\mathrm{D}_{\mathbf{r}_1^\mathcal{W%
},j^{\prime }j}^\beta(\{p|\mathbf{R}\})\phi_{\mathbf{r}_1^\mathcal{W}%
,j^{\prime }}^\beta$. Note that $\phi_{\mathbf{r}_1^\mathcal{W},j}^\beta$
may not be localized on $\mathbf{r}_1^{\mathcal{W}}$. Then we can obtain
other real-space orbitals: $\{p_J|\mathbf{R}_J+\mathbf{R}\}\phi_{\mathbf{r}%
_1^\mathcal{W},j}^\beta$, which constitute a complete basis for the $%
\mathcal{SG}$. Through Bloch summation:
\begin{equation}  \label{app:blochsum}
\psi_{\mathbf{k},J,j}^\beta=\sum_{\mathbf{R}} e^{i\mathbf{k}\cdot(\mathbf{r}%
_{J}^\mathcal{W}+\mathbf{R})}\{p_J|\mathbf{R}_J\}\phi_{\mathbf{r}_1^\mathcal{%
W},j}^\beta,
\end{equation}
which is the basis for some $\mathbf{k}$ point in the BZ.  By the technique shown in Sec. \ref{red}, one can obtain all the ${n_{\mathbf{k}}^{\alpha}}'s$ for the AI corresponding to this Wyckoff position and site-symmetry irrep.
\begin{table}[htbp!]
\centering
\begin{tabular}{|c|c|}
\hline
$C_1$ & $E$ \\ \hline
$D_s^1$ & 2 \\ \hline
\end{tabular}%
\caption{The irrep $\mathcal{D}_s$ for $C_1$ considering $\mathcal{T}$. Note
we only consider the doubled-valued irreps, thus the character would be
multiplied by $-1$ when considering a $2\protect\pi$ more rotation.}
\label{app:c1}
\end{table}
\begin{table}[htbp!]
\centering
\begin{tabular}{|c|c|c|}
\hline
$C_i$ & $E$ & $P$ \\ \hline
$D_s^1$ & 2 & $2$ \\ \hline
$D_s^2$ & 2 & $-2$ \\ \hline
\end{tabular}%
\caption{The irrep $\mathcal{D}_s$ for $C_i$ considering $\mathcal{T}$.}
\label{app:ci}
\end{table}

\emph{\textbf{Step 3}}:
\textit{Make Smith normal decomposition.} According to the
Wyckoff positions listed in Table \ref{app:sg2wyck}, and using the
corresponding site-group irreps shown in Tables \ref{app:c1} and \ref{app:ci}%
, we thus obtain 17 AIs, denoted by $\mathrm{n}_{m\mathcal{W}}^i$ with the
superscript labeling the site-group-irrep in the same order shown in Tables %
\ref{app:c1} and \ref{app:ci}, and the subscript labels the Wyckoff
position. They are listed in Table \ref{app:sg2ai}. These 17 vectors are
generally not independent with each other, we can make a so-called Smith
normal decomposition which will give $d_{\mathrm{AI}}(\le17)$ AI basis
vectors through the linear combination of these 17 AI vectors. The results
are shown in Table \ref{app:sg2aibasis} where the 9 basis vectors are
printed explicitly and they are in an ascending order of the common factors. So $d_{\mathrm{AI}}^{\mathcal{SG}\mathbf{2}}=9$ which is equal to $d_{\mathrm{BS}}^{\mathcal{SG}\mathbf{2}}$. Ref.\ \onlinecite{Po} proved that it  holds for each of  230 $\mathcal{SG}s$ \cite{Po}.
It is easy to find that $\mathbf{a}_1,\ldots,\mathbf{a}_5$ have no
common factors while $\mathbf{a}_6$,$\mathbf{a}_7$,$\mathbf{a}_8$ have a
common factor (=2), and $\mathbf{a}_9$ have a common factor (=4). Thus the
symmetry indicator (SI) group is $X_{\mathrm{BS}}=\mathbb{Z}_2\times\mathbb{Z%
}_2\times\mathbb{Z}_2\times\mathbb{Z}_4$. Further more we can find that the
condition for the filling of a band insulator is $2\mathbb{N}$.

\begin{table}[tbp]
\centering
\begin{tabular}{|c|c|c|c|c|c|c|c|c|c|c|c|c|c|c|c|c|c|}
\hline
$\mathcal{SG}2$ & $\mathrm{n}_{2i}^{1}$ & $\mathrm{n}_{1h}^{1}$ & $\mathrm{n}%
_{1h}^{2}$ & $\mathrm{n}_{1g}^{1}$ & $\mathrm{n}_{1g}^{2}$ & $\mathrm{n}%
_{1f}^{1}$ & $\mathrm{n}_{1f}^{2}$ & $\mathrm{n}_{1e}^{1}$ & $\mathrm{n}%
_{1e}^{2}$ & $\mathrm{n}_{1d}^{1}$ & $\mathrm{n}_{1d}^{2}$ & $\mathrm{n}%
_{1c}^{1}$ & $\mathrm{n}_{1c}^{2}$ & $\mathrm{n}_{1b}^{1}$ & $\mathrm{n}%
_{1b}^{2}$ & $\mathrm{n}_{1a}^{1}$ & $\mathrm{n}_{1a}^{2}$ \\ \hline
$\nu$ & 4 & 2 & 2 & 2 & 2 & 2 & 2 & 2 & 2 & 2 & 2 & 2 & 2 & 2 & 2 & 2 & 2 \\
${\color{red} \Gamma}_1^1$ & 1 & 1 & 0 & 1 & 0 & 1 & 0 & 1 & 0 & 1 & 0 & 1 &
0 & 1 & 0 & 1 & 0 \\
${\color{red} \Gamma}_2^1$ & 1 & 0 & 1 & 0 & 1 & 0 & 1 & 0 & 1 & 0 & 1 & 0 &
1 & 0 & 1 & 0 & 1 \\
${\color{red} X}_1^1$ & 1 & 1 & 0 & 0 & 1 & 1 & 0 & 0 & 1 & 0 & 1 & 0 & 1 & 1
& 0 & 1 & 0 \\
${\color{red} X}_2^1$ & 1 & 0 & 1 & 1 & 0 & 0 & 1 & 1 & 0 & 1 & 0 & 1 & 0 & 0
& 1 & 0 & 1 \\
${\color{red} Y}_1^1$ & 1 & 1 & 0 & 0 & 1 & 0 & 1 & 1 & 0 & 0 & 1 & 1 & 0 & 0
& 1 & 1 & 0 \\
${\color{red} Y}_2^1$ & 1 & 0 & 1 & 1 & 0 & 1 & 0 & 0 & 1 & 1 & 0 & 0 & 1 & 1
& 0 & 0 & 1 \\
${\color{red} Z}_1^1$ & 1 & 1 & 0 & 0 & 1 & 0 & 1 & 0 & 1 & 1 & 0 & 1 & 0 & 1
& 0 & 0 & 1 \\
${\color{red} Z}_2^1$ & 1 & 0 & 1 & 1 & 0 & 1 & 0 & 1 & 0 & 0 & 1 & 0 & 1 & 0
& 1 & 1 & 0 \\
${\color{red} U}_1^1$ & 1 & 1 & 0 & 1 & 0 & 0 & 1 & 0 & 1 & 1 & 0 & 0 & 1 & 0
& 1 & 1 & 0 \\
${\color{red} U}_2^1$ & 1 & 0 & 1 & 0 & 1 & 1 & 0 & 1 & 0 & 0 & 1 & 1 & 0 & 1
& 0 & 0 & 1 \\
${\color{red} T}_1^1$ & 1 & 1 & 0 & 1 & 0 & 1 & 0 & 0 & 1 & 0 & 1 & 1 & 0 & 0
& 1 & 0 & 1 \\
${\color{red} T}_2^1$ & 1 & 0 & 1 & 0 & 1 & 0 & 1 & 1 & 0 & 1 & 0 & 0 & 1 & 1
& 0 & 1 & 0 \\
${\color{red} S}_1^1$ & 1 & 1 & 0 & 1 & 0 & 0 & 1 & 1 & 0 & 0 & 1 & 0 & 1 & 1
& 0 & 0 & 1 \\
${\color{red} S}_2^1$ & 1 & 0 & 1 & 0 & 1 & 1 & 0 & 0 & 1 & 1 & 0 & 1 & 0 & 0
& 1 & 1 & 0 \\
${\color{red} R}_1^1$ & 1 & 1 & 0 & 0 & 1 & 1 & 0 & 1 & 0 & 1 & 0 & 0 & 1 & 0
& 1 & 0 & 1 \\
${\color{red} R}_2^1$ & 1 & 0 & 1 & 1 & 0 & 0 & 1 & 0 & 1 & 0 & 1 & 1 & 0 & 1
& 0 & 1 & 0 \\ \hline
\end{tabular}%
\caption{The 17 AI vectors for $\mathcal{SG}2$.}
\label{app:sg2ai}
\end{table}
\section{Reduction of AI on HSPs}\label{red}
Focused on HSPs,
we can then easily obtain the number of occurrences for the irrep $\alpha$ of
$\mathcal{G}(\mathbf{k})$: $n_{\mathbf{k}}^\alpha$ for the above basis
(generally reducible) by:
\begin{equation}  \label{app:n}
n_{\mathbf{k}}^\alpha=\frac{1}{|\mathcal{G}(\mathbf{k})|}\sum_{g\in\mathcal{G%
}(\mathbf{k})}{\chi_\mathbf{k}^\alpha(g)}^*\chi_\mathbf{k}(g),
\end{equation}
where $\chi$ denotes the character. $\chi_\mathbf{k}(g)$ is the character
for the basis in Eq. (\ref{app:blochsum}), can be obtained through ($g\in%
\mathcal{G}(\mathbf{k})$):
\begin{equation}  \label{app:DD}
g \psi_{\mathbf{k},J,j}^\beta=\sum_{J^{\prime },j^{\prime }}\mathrm{D}%
_{J^{\prime },j^{\prime };J,j}(g)\psi_{\mathbf{k},J^{\prime },j^{\prime
}}^\beta,
\end{equation}
where $\chi_\mathbf{k}=tr(\mathrm{D})$. The $\mathrm{D}$ is obtained by
first knowing the permutation (i.e. $J\rightarrow J^{\prime }$) of the
atoms:
\begin{equation}  \label{app:JJ'}
g\{p_J|\mathbf{R}_J\}=\{p_{J^{\prime }}|\mathbf{R}_{J^{\prime }}+\mathbf{R}%
\}g^{\prime },
\end{equation}
where $g^{\prime }\in G(\mathbf{r}^\mathcal{W}_1)$, thus,
\begin{equation}  \label{app:D}
\mathrm{D}_{J^{\prime },j^{\prime };J,j}(g)=e^{i\mathbf{k}\cdot(\mathbf{r}_J^%
\mathcal{W}-\mathbf{r}_{J^{\prime }}^\mathcal{W})-i\mathbf{k}\cdot\mathbf{R}}%
\mathrm{D}_{\mathbf{r}_1^\mathcal{W},j^{\prime }j}^\beta(g^{\prime }).
\end{equation}
Note that when we only need to calculate the trace of the representation
matrix $\mathrm{D}(g)$, we can just consider $J=J^{\prime }$ but for some
cases we should know exactly the representation matrix, e.g. when
calculating the mirror Chern number, we should know the representation
matrix for the mirror operation.
\section{Atomic basis vectors}\label{AIS}

In this section, we list the AI basis vectors for all the $\mathcal{SG}s$ we
encounter in this work.
\begin{table}[htbp!]
\subtable[The 9 AI basis vectors for $\mathcal{SG}\mathbf{2}$. Here $\nu$ is the number of the bands. It is also called the filling number. Staring from the 3rd row, we give the number $n_{\mathbf{k}}^\alpha$ in order. We omit the notation $n$ for clarity:  The first column of these rows  gives the information of the HSP and its irrep completely.]{
  \centering
  \begin{tabular}{|c|c|c|c|c|c|c|c|c|c|}
    \hline
     $\mathcal{SG}\mathbf{2}$ & $\mathbf{a}_1$ & $\mathbf{a}_2$ & $\mathbf{a}_3$ & $\mathbf{a}_4$ & $\mathbf{a}_5$ & $\mathbf{a}_6$ & $\mathbf{a}_7$ & $\mathbf{a}_8$ &$\mathbf{a}_9$ \\\hline
  $\nu$ & 4 & 2 & 2 & 2 & 2 &  4 &  4 &  4 & 8 \\
  ${\color{red} \Gamma}_1^1$ & 1 & 1 & 1 & 1 & 1 &  2 &  2 &  2 & 4  \\
 ${\color{red} \Gamma}_2^1$ &  1 & 0 & 0 & 0 & 0 & 0 & 0 & 0 & 0  \\
  ${\color{red} X}_1^1$ & 1 & 1 & 0 & 1 & 0 &  2 &  2 & 0 & 4  \\
  ${\color{red} X}_2^1$ & 1 & 0 & 1 & 0 & 1 & 0 & 0 &  2 & 0 \\
  ${\color{red} Y}_1^1$ & 1 & 1 & 0 & 0 & 1 &  2 & 0 &  2 & 4  \\
  ${\color{red} Y}_2^1$ & 1 & 0 & 1 & 1 & 0 & 0 & 2 & 0 & 0  \\
  ${\color{red} Z}_1^1$ & 1 & 1 & 0 & 0 & 0 & 0 & 0 & 0 & 0  \\
  ${\color{red} Z}_2^1$ & 1 & 0 & 1 & 1 & 1 &  2 &  2 &  2 & 4 \\
  ${\color{red} U}_1^1$ & 1 & 1 & 1 & 0 & 0 & 0 &  2 &  2 & 4  \\
  ${\color{red} U}_2^1$ & 1 & 0 & 0 & 1 & 1 &  2 & 0 & 0 & 0 \\
  ${\color{red} T}_1^1$ & 1 & 1 & 1 & 1 & 0 &  2 &  2 &  2 & 4  \\
  ${\color{red} T}_2^1$ & 1 & 0 & 0 & 0 & 1 & 0 & 0 & 0 & 0 \\
  ${\color{red} S}_1^1$ & 1 & 1 & 1 & 0 & 1 & 2 & 2 &  2 & 4  \\
  ${\color{red} S}_2^1$ & 1 & 0 & 0 & 1 & 0 & 0 & 0 & 0 & 0  \\
  ${\color{red} R}_1^1$ & 1 & 1 & 0 & 1 & 1 & 2 & 2 & 2 & 4  \\
  ${\color{red} R}_2^1$ & 1 & 0 & 1 & 0 & 0 & 0 & 0 & 0 & 0  \\

    \hline
  \end{tabular}\label{app:sg2aibasis}} \qquad \centering
\subtable[The 5 AI basis vectors for $\mathcal{SG}\mathbf{11}$.]{
    \centering
  \begin{tabular}{|c|c|c|c|c|c|}
    \hline
 $\mathcal{SG}\mathbf{11}$& $\mathbf{a_1}$& $\mathbf{a_2}$&$\mathbf{a_3}$&$\mathbf{a_4}$&$\mathbf{a_5}$\\
 \hline
 $\nu$&4 & 4 & 0 & 0 & 0 \\
  $\Gamma_1^1$&1 & 2 & 0 & 0 & 0 \\
  $\Gamma_2^1$&1 & 2 & 0 & 0 & 0 \\
  $\Gamma_3^1$&1 & 0 & 0 & 0 & 0 \\
  $\Gamma_4^1$&1 & 0 & 0 & 0 & 0 \\
  $B_1^1$&1 & 0 & 0 & 2 & 0 \\
 $B_2^1$& 1 & 0 & 0 & 2 & 0 \\
  $B_3^1$&1 & 2 & 0 & -2 & 0 \\
 $B_4^1$&1 & 2 & 0 & -2 & 0 \\
  $Y_1^1$&1 & 0 & 2 & 0 & 0 \\
  $Y_2^1$&1 & 0 & 2 & 0 & 0 \\
  $Y_3^1$&1 & 2 & -2 & 0 & 0 \\
  $Y_4^2$&1 & 2 & -2 & 0 & 0 \\
  ${\color{red} Z}_1^2$&1 & 1 & 0 & 0 & 0 \\
   ${\color{red} C}_1^2$&1 & 1 & 0 & 0 & 0 \\
 ${\color{red}D}_1^2$&1 & 1 & 0 & 0 & 0 \\
 $A_1^1$&1 & 2 & -2 & -2 & 4 \\
 $A_2^1$&1 & 2 & -2 & -2 & 4 \\
 $A_3^1$&1 & 0 & 2 & 2 & -4 \\
 $A_4^1$&1 & 0 & 2 & 2 & -4 \\
 ${\color{red}E}_1^2$& 1& 1 & 0 & 0 & 0 \\

    \hline
  \end{tabular}\label{app:sg11aibasis}} \qquad
\subtable[The 7 AI basis vectors for $\mathcal{SG}\mathbf{12}$.]{
   \centering
  \begin{tabular}{|c|c|c|c|c|c|c|c|}
    \hline
 $\mathcal{SG}\mathbf{12}$& $\mathbf{a_1}$& $\mathbf{a_2}$&$\mathbf{a_3}$&$\mathbf{a_4}$&$\mathbf{a_5}$& $\mathbf{a_6}$& $\mathbf{a}_7$\\
 \hline

 $\nu$& 4 & 4 & -8 & 2 & -4 & 8 & -8 \\
 $\Gamma_1^1$&1 & 2 & -4 & 1 & -2 & 4 & -4 \\
   $\Gamma_2^1$&1 & 2 & -4 & 1 & -2 & 4 & -4 \\
   $\Gamma_3^1$&1 & 0 & 0 & 0 & 0 & 0 & 0 \\
   $\Gamma_4^1$&1 & 0 & 0 & 0 & 0 & 0 & 0 \\
   $A_1^1$&1 & 0 & -1 & 0 & 0 & 2 & -4 \\
   $A_2^1$&1 & 0 & -1 & 0 & 0 & 2 & -4 \\
   $A_3^1$&1 & 2 & -3 & 1 & -2 & 2 & 0 \\
   $A_4^1$&1 & 2 & -3 & 1 & -2 & 2 & 0 \\
   $Z_1^1$&1 & 0 & -2 & 1 & 0 & 2 & -4 \\
   $Z_2^1$&1 & 0 & -2 & 1 & 0 & 2 & -4 \\
   $Z_3^1$&1 & 2 & -2 & 0 & -2 & 2 & 0 \\
   $Z_4^1$&1 & 2 & -2 & 0 & -2 & 2 & 0 \\
   $M_1^1$&1 & 2 & -3 & 0 & -2 & 4 & 0 \\
   $M_2^1$&1 & 2 & -3 & 0 & -2 & 4 & 0 \\
   $M_3^1$&1 & 0 & -1 & 1 & 0 & 0 & -4 \\
   $M_4^1$&1 & 0 & -1 & 1 & 0 & 0 & -4 \\
   ${\color{red}L}_1^1$&1 & 1 & -1 & 1 & -2 & 0 & 0 \\
${\color{red}L}_2^1$& 1 & 1 & -3 & 0 & 0 & 4 & -4 \\
 ${\color{red}V}_1^1$&1 & 1 & 0 & 0 & 0 & 0 & 0 \\
 ${\color{red}V}_2^1$&1 & 1 & -4 & 1 & -2 & 4 & -4 \\
    \hline
  \end{tabular}\label{app:sg12aibasis} }
\par
\qquad
\end{table}

\begin{table}[htbp!]
\subtable[The 3 AI basis vectors for $\mathcal{SG}\mathbf{61}$.]{
  \centering
  \begin{tabular}{|c|c|c|c|}
    \hline
 $\mathcal{SG}\mathbf{61}$& $\mathbf{a_1}$& $\mathbf{a_2}$&$\mathbf{a_3}$\\
 \hline

 $\nu$&16 & 8 & -32 \\
 $\Gamma_1^2$&4 & 4 & -12 \\
$\Gamma_2^2$ &4 & 0 & -4 \\
  $Y_1^2$&4 & 2 & -8 \\
  $Y_2^2$&4 & 2 & -8 \\
  $X_1^2$&4 & 2 & -8 \\
  $X_2^2$&4 & 2 & -8 \\
  $Z_1^2$&4 & 2 & -8 \\
  $Z_2^2$&4 & 2 & -8 \\
  ${\color{red}U}_1^2$&2 & 1 & -4 \\
 ${\color{red}U}_2^2$&2 & 1 & -4 \\
 ${\color{red}T}_1^2$&2 & 1 & -4 \\
${\color{red}T}_2^2$ &2 & 1 & -4 \\
 ${\color{red}S}_1^2$&2 & 1 & -4 \\
 ${\color{red}S}_2^2$&2 & 1 & -4 \\
 ${\color{red}R}_1^1$&1 & 0 & 0 \\
 ${\color{red}R}_2^1$&1 & 0 & 0 \\
 ${\color{red}R}_3^1$&1 & 0 & 0 \\
 ${\color{red}R}_4^1$&1 & 0 & 0 \\
 ${\color{red}R}_5^1$&1 & 1 & -4 \\
 ${\color{red}R}_6^1$&1 & 1 & -4 \\
 ${\color{red}R}_7^1$&1 & 1 & -4 \\
 ${\color{red}R}_8^1$&1 & 1 & -4 \\
    \hline
  \end{tabular}
\label{app:sg61aibasis}} \qquad \centering

\par
\qquad \centering
\par
\subtable[The 8 AI basis vectors for $\mathcal{SG}\mathbf{136}$. ]{
  \centering
  \begin{tabular}{|c|c|c|c|c|c|}
    \hline
 $\mathcal{SG}\mathbf{136}$& $\mathbf{a_1}$& $\mathbf{a_2}$&$\mathbf{a_3}$&$\mathbf{a_4}$&$\mathbf{a_5}$\\
 \hline

 $\nu$&8 & 8 & 8 & 4 & -48 \\
 $\Gamma_1^2$&1 & 2 & 2 & 1 & -12 \\
 $\Gamma_2^2$&1 & 0 & 2 & 1 & -8 \\
 $\Gamma_3^2$&1 & 0 & 0 & 0 & 0 \\
 $\Gamma_4^2$&1 & 2 & 0 & 0 & -4 \\
 $K_1^2$&1 & 1 & 0 & 1 & -4 \\
 $K_2^2$&1 & 1 & 0 & 1 & -4 \\
 $K_3^2$&1 & 1 & 2 & 0 & -8 \\
 $K_4^2$&1 & 1 & 2 & 0 & -8 \\
 $K_1^4$&2 & 2 & 2 & 1 & -12 \\
 $K_1^4$&2 & 2 & 2 & 1 & -12 \\
 $K_1^1$&1 & 1 & 1 & 0 & -4 \\
 $K_2^1$&1 & 1 & 1 & 0 & -4 \\
 $K_3^1$&1 & 1 & 1 & 0 & -4 \\
 $K_4^1$&1 & 1 & 1 & 0 & -4 \\
 $K_5^1$&1 & 1 & 1 & 1 & -8 \\
 $K_6^1$&1 & 1 & 1 & 1 & -8 \\
 $K_7^1$&1 & 1 & 1 & 1 & -8 \\
 $K_8^1$&1 & 1 & 1 & 1 & -8 \\
 $K_1^2$&2 & 2 & 2 & 1 & -12 \\
 $K_2^2$&2 & 2 & 2 & 1 & -12 \\

    \hline
  \end{tabular}\label{app:sg227aibasis}} \qquad \centering
\qquad

\end{table}

\begin{table}[htbp!]

\subtable[The 8 AI basis vectors for $\mathcal{SG}\mathbf{166}$.]{
  \centering
  \begin{tabular}{|c|c|c|c|c|c|c|c|c|}
    \hline
 $\mathcal{SG}\mathbf{166}$& $\mathbf{a_1}$& $\mathbf{a_2}$&$\mathbf{a_3}$&$\mathbf{a_4}$&$\mathbf{a_5}$& $\mathbf{a_6}$&$\mathbf{a_7}$&$\mathbf{a_8}$\\
 \hline

 $\nu$&12 & 6 & 6 & -4 & 0 & 2 & -4 & 8 \\
 $\Gamma_1^1$&1 & 1 & 1 & 0 & 0 & 0 & 0 & 0 \\
 $\Gamma_2^1$&1 & 1 & 1 & 0 & 0 & 0 & 0 & 0 \\
  $\Gamma_3^2$&2 & 2 & 2 & -1 & 0 & 1 & -2 & 4 \\
  $\Gamma_4^1$&1 & 0 & 0 & 0 & 0 & 0 & 0 & 0 \\
   $\Gamma_5^1$&1 & 0 & 0 & 0 & 0 & 0 & 0 & 0 \\
   $\Gamma_6^2$&2 & 0 & 0 & -1 & 0 & 0 & 0 & 0 \\
   $Z_1^1$&1 & 1 & 0 & 0 & 0 & 0 & 0 & 0 \\
   $Z_2^1$&1 & 1 & 0 & 0 & 0 & 0 & 0 & 0 \\
   $Z_3^2$&2 & 2 & 0 & -1 & 1 & 0 & -2 & 4 \\
   $Z_4^1$&1 & 0 & 1 & 0 & 0 & 0 & 0 & 0 \\
   $Z_5^1$&1 & 0 & 1 & 0 & 0 & 0 & 0 & 0 \\
   $Z_6^2$&2 & 0 & 2 & -1 & -1 & 1 & 0 & 0 \\
   $L_1^1$&3 & 1 & 2 & -1 & 1 & 0 & 0 & 4 \\
   $L_2^1$&3 & 1 & 2 & -1 & 1 & 0 & 0 & 4 \\
   $L_3^1$&3 & 2 & 1 & -1 & -1 & 1 & -2 & 0 \\
   $L_4^1$&3 & 2 & 1 & -1 & -1 & 1 & -2 & 0 \\
   $F_1^1$&3 & 1 & 1 & -1 & 0 & 1 & 0 & 0 \\
   $F_2^1$&3 & 1 & 1 & -1 & 0 & 1 & 0 & 0 \\
   $F_3^1$&3 & 2 & 2 & -1 & 0 & 0 & -2 & 4 \\
   $F_4^1$&3 & 2 & 2 & -1 & 0 & 0 & -2 & 4 \\

    \hline
  \end{tabular}
 \label{app:sg166aibasis}} \qquad\qquad
\end{table}

\begin{table}[htbp!]

\subtable[The 3 AI basis vectors for $\mathcal{SG}\mathbf{216}$.]{
  \centering
  \begin{tabular}{|c|c|c|c|c|c|c|}
    \hline
 $\mathcal{SG}\mathbf{216}$& $\mathbf{a_1}$& $\mathbf{a_2}$&$\mathbf{a_3}$&$\mathbf{a_4}$&$\mathbf{a_5}$&$\mathbf{a_6}$\\
 \hline

$\nu$& 12 & 4 & 2 & 2 & 2 & -4  \\
 $\Gamma_1^2$&1 & 0 & 1 & 1 & 1 & -2  \\
 $\Gamma_2^2$& 1 & 0 & 0 & 0 & 0 & 0  \\
  $\Gamma_3^4$&2 & 1 & 0 & 0 & 0 & 0  \\
  $X_1^2$&3 & 1 & 1 & 1 & 0 & -2  \\
  $X_2^2$&3 & 1 & 0 & 0 & 1 & 0  \\
  $L_1^1$&2 & 1 & 0 & 0 & 0 & 0  \\
  $L_2^1$&2 & 1 & 0 & 0 & 0 & 0  \\
  $L_3^2$&4 & 1 & 1 & 1 & 1 & -2  \\
  $W_1^1$&3 & 1 & 0 & 1 & 1 & -2  \\
  $W_2^1$&3 & 1 & 0 & 1 & 0 & 0  \\
  $W_3^1$&3 & 1 & 1 & 0 & 0 & 0  \\
  $W_4^1$&3 & 1 & 1 & 0 & 1 & -2  \\

    \hline
  \end{tabular}
\label{app:sg216aibasis}}
\qquad
\par
 \qquad
\subtable[The 8 AI basis vectors for $\mathcal{SG}\mathbf{227}$. ]{
  \centering
  \begin{tabular}{|c|c|c|c|c|c|c|c|c|}
    \hline
 $\mathcal{SG}\mathbf{227}$& $\mathbf{a_1}$& $\mathbf{a_2}$&$\mathbf{a_3}$&$\mathbf{a_4}$&$\mathbf{a_5}$& $\mathbf{a_6}$&$\mathbf{a_7}$&$\mathbf{a_8}$\\
 \hline

$\nu$& 24 & 8 & 8 & 8 & -24 & 8 & 4 & -48 \\
 $\Gamma_1^2$&1 & 0 & 0 & 1 & -2 & 1 & 1 & -4 \\
  $\Gamma_2^2$&1 & 0 & 0 & 1 & -2 & 1 & 0 & -4 \\
  $\Gamma_3^4$&2 & 1 & 2 & 1 & -4 & 1 & 0 & -8 \\
  $\Gamma_4^2$&1 & 0 & 0 & 0 & 0 & 0 & 0 & 0 \\
  $\Gamma_5^2$&1 & 0 & 0 & 0 & 0 & 0 & 1 & 0 \\
  $\Gamma_6^4$&2 & 1 & 0 & 0 & 0 & 0 & 0 & 0 \\
  $X_1^4$&6 & 2 & 2 & 2 & -6 & 2 & 1 & -12 \\
  $L_1^1$&2 & 1 & 1 & 1 & -2 & 0 & 0 & -4 \\
  $L_2^1$&2 & 1 & 1 & 1 & -2 & 0 & 0 & -4 \\
  $L_3^2$&4 & 1 & 2 & 2 & -5 & 1 & 1 & -8 \\
  $L_4^1$&2 & 1 & 1 & 0 & -2 & 1 & 0 & -4 \\
  $L_5^1$&2 & 1 & 1 & 0 & -2 & 1 & 0 & -4 \\
  $L_6^2$&4 & 1 & 0 & 1 & -3 & 2 & 1 & -8 \\
  $W_1^1$&3 & 1 & 1 & 1 & -2 & 1 & 0 & -4 \\
  $W_2^1$&3 & 1 & 1 & 1 & -2 & 1 & 0 & -4 \\
  $W_3^1$&3 & 1 & 1 & 1 & -4 & 1 & 1 & -8 \\
  $W_4^1$&3 & 1 & 1 & 1 & -4 & 1 & 1 & -8 \\
  $W_5^2$&6 & 2 & 2 & 2 & -6 & 2 & 1 & -12 \\
    \hline
  \end{tabular}\label{app:sg227aibasis}}
\end{table}


\bibliography{refs}